\documentclass[11pt,a4paper]{article}
\usepackage{jheppub}
\usepackage{graphicx}
\usepackage{caption,subcaption}
\usepackage[linesnumbered,ruled,vlined]{algorithm2e}
\DontPrintSemicolon

\newcommand{\be}{\begin{equation}}
\newcommand{\beq}{\begin{equation}}
\newcommand{\bi}{\begin{itemize}}
\newcommand{\ei}{\end{itemize}}
\newcommand{\en}{\end{equation}}
\newcommand{\eeq}{\end{equation}}
\newcommand{\bea}{\begin{eqnarray}}
\newcommand{\ena}{\end{eqnarray}}
\newcommand{\hbo}{\hbox to 1 true cm {\hfill } }

\newcommand{\dlangle}{\left\langle \kern-.17em \left\langle}
\newcommand{\drangle}{\right\rangle \kern-.17em \right\rangle}
\newcommand{\dd}{{\rm d}}
\newcommand{\EMAX}{E_{\mathrm{max}}}
\newcommand{\EMIN}{E_{\mathrm{min}}}
\newcommand{\NSWEEPS}{N_{\mathrm{SW}}}
\newcommand{\NTHERM}{N_{\mathrm{TH}}}
\newcommand{\NAVG}{N_{\mathrm{A}}}
\newcommand{\NRM}{N_{\mathrm{RM}}}
\newcommand{\DE}{\delta _E}
\newcommand{\e}{\mathrm{e}}

\title{
An efficient algorithm for numerical computations of continuous densities of states
}

\author[a]{K. Langfeld}
\author[b]{B. Lucini}
\author[c]{R. Pellegrini}
\author[a]{A. Rago}

\affiliation[a]{
Centre for Mathematical Sciences, Plymouth University, 
Plymouth, PL4 8AA, UK 
}
\affiliation[b]{
  College of Science, Swansea University,
  Singleton Park, Swansea SA2 8PP, UK
}

\affiliation[c]{School of Physics and Astronomy, University of 
Edinburgh, Edinburgh EH9 3FD, UK}

\emailAdd{kurt.langfeld@plymouth.ac.uk}
\emailAdd{b.lucini@swansea.ac.uk}
\emailAdd{r.pellegrini@ed.ac.uk}
\emailAdd{antonio.rago@plymouth.ac.uk}

\abstract{
In Wang-Landau type algorithms, Monte-Carlo updates are performed with
respect to the density of states, which is iteratively refined during
simulations. The partition function and thermodynamic observables are
then obtained by standard integration. In this work, our recently
introduced method in this class (the LLR approach) is analysed and
further developed. Our approach is a histogram free method
particularly suited for systems with continuous degrees of freedom
giving rise to a continuum density of states, as it is commonly found
in Lattice Gauge Theories and in some Statistical Mechanics systems. We show that
the method possesses an exponential error suppression that allows us to
estimate the density of states over several orders of magnitude with nearly-constant
{\it relative} precision. We explain how ergodicity issues can  be avoided
and how expectation values of arbitrary observables can be obtained
within this framework. We then demonstrate the method using Compact U(1)
Lattice Gauge Theory as a show case. A thorough study of the
algorithm parameter dependence of the results is performed and 
compared with the analytically expected behaviour. We obtain high
precision values for the critical coupling for the phase 
transition and for the peak value of the specific heat for lattice sizes
ranging from $8^4$ to $20^4$. Our results perfectly agree with the
reference values reported in the literature, which covers lattice
sizes up to $18^4$. Robust results for the $20^4$ volume are obtained for
the first time. This latter investigation, which, due to strong metastabilities
developed at the pseudo-critical coupling of the system, so far has been out of reach even on
supercomputers with importance sampling approaches, has been
performed to high accuracy with modest computational resources. This
shows the potential of the method for studies of first order phase
transitions. Other situations where the method is expected to be
superior to importance sampling techniques are pointed out.
}

\begin{document}
\setlength{\parindent}{0pt}

\maketitle
\bibliographystyle{JHEP}

\section{Introduction and motivations}
Monte-Carlo methods are widely used in Theoretical Physics,
Statistical Mechanics and Condensed Matter (for an overview, see
e.g.~\cite{landau2014guide}).  Since the inception of the
field~\cite{metropolis}, most of the applications have relied on importance
sampling, which allows us to evaluate stochastically with a
controllable error  multi-dimensional integrals of localised
functions. These methods have immediate applications when one needs to
compute thermodynamic properties, since statistical averages of (most)
observables can be computed efficiently with importance sampling
techniques. Similarly, in Lattice Gauge Theories, most quantities of
interest can be expressed in the path integral formalism as ensemble averages
over a positive-definite (and sharply peaked) measure, which, once
again, provide an ideal scenario for applying importance sampling
methods. 

However, there are noticeable cases in which Monte-Carlo importance
sampling methods are either very inefficient or produce inherently wrong results for well
understood reasons. Among those cases, some of the most relevant
situations include systems with a sign problem (see~\cite{Aarts:2015kea} for a recent
review), direct computations of free energies (comprising
the study of properties of interfaces), systems with strong
metastabilities (for instance, a system with a first order phase
transition in the region in which the phases coexist) and systems with
a rough free energy landscape. Alternatives to importance sampling
techniques do exist, but generally they are less efficient in
standard cases and hence their use is limited to ad-hoc situations
in which more standard methods are inapplicable. Noticeable
exceptions are micro-canonical methods, which have experienced a surge in
interest in the past fifteen years. Most of the growing popularity of
those methods is due to the work of Wang and
Landau~\cite{Wang:2001ab}, which provided an efficient algorithm to
access the density of states in a statistical system with a discrete
spectrum. Once the density of states is known, the partition function
(and from it all thermodynamic properties of the system) can be
reconstructed by performing one-dimensional numerical integrals. The
generalisation of the Wang-Landau algorithm to systems with a continuum spectrum is far from
straightforward~\cite{Xu:2007aa,Sinha:2007aa}. To overcome this
limitation, a very promising method, here referred to as the
Logarithmic Linear Relaxation (LLR) algorithm, was introduced
in~\cite{Langfeld:2012ah}. The potentialities of the  method were
demonstrated in subsequent studies of systems afflicted by a sign
problem~\cite{Langfeld:2014nta,Gattringer:2015lra}, in the computation
of the Polyakov loop probability distribution function in two-colour
QCD with heavy quarks at finite density~\cite{Langfeld:2013xbf} and --
rather unexpectedly -- even in the determination of thermodynamic
properties of systems with a discrete energy
spectrum~\cite{Guagnelli:2012dk}.   

The main purpose of this work is to discuss in detail some improvements of the
original LLR algorithm and to formally prove that expectation values
of observables computed with this method converge to the correct
result, which fills  a gap in the current literature. In addition, we apply the algorithm to
the study of Compact U(1) Lattice Gauge Theory, a system with severe
metastabilities at its first order phase transition point that make the
determination of observables near the transition very difficult
from a numerical point of view. We find that in the LLR approach
correlation times near criticality grow {\em at most quadratically} with the
volume, as opposed to the exponential growth that one expects with
importance sampling methods. This investigation shows the efficiency
of the LLR method when dealing with systems having a first order phase
transition. These results suggest that the LLR method can be efficient
at overcoming numerical metastabilities in other classes of systems with a multi-peaked
probability distribution, such as those with rough free energy
landscapes (as commonly found, for instance, in models of protein 
folding or spin glasses).

The rest of the paper is organised as follows. In
Sect.~\ref{sect:density} we cover the formal general aspects of the
algorithm. The investigation of Compact U(1) Lattice Gauge Theory is
reported in Sect.~\ref{sect:u1}. A critical analysis of our
findings, our conclusions and our future plans are presented in
Sect.~\ref{sect:conclusions}. Finally, some technical material is
discussed in the appendix. Some preliminary results of this study
have already been presented in~\cite{Pellegrini:2014gha}.  

\section{Numerical determination of the density of states}
\label{sect:density}

\subsection{The density of states}
Owing to formal similarities between the two fields, the approach we
are proposing can be applied to both Statistical Mechanics and Lattice
Field Theory systems. In order to keep
the discussion as general as possible, we shall introduce notations
and conventions that can describe simultaneously both cases. We shall
consider a system described by the set of dynamical variables
$\phi$, which could represent a set of spin or field variables and are
assumed to be continuous. The action (in the field theory
case) or the Hamiltonian (for the statistical system) is indicated by
$S$ and the coupling (or inverse temperature) by $\beta$. Since the
product $\beta S$ is dimensionless, without loss of generality we will
take both $S$ and $\beta$ dimensionless. 

We consider a system with a finite volume $V$, which will be sent to
infinity in the final step of our calculations. The finiteness of $V$
in the intermediate steps allows us to define naturally a measure
over the variables $\phi$, which we shall call ${\cal D}
  \phi$. Properties of the system can be derived from the function   
\[
Z(\beta)=\int {\cal D} \phi \;  \e^{\beta S[\phi]}\,.
\]
which defines the canonical partition function for the statistical
system or  the path integral in the field theory case. The density of
state (which is a function of the value of $S[\phi] = E$) is formally
defined by the integral 
\be 
\rho(E) = \int {\cal D} \phi \;  \delta \Bigl( S[\phi]-E \Bigr) \, .
\label{eq:den}
\en 
In terms of $\rho(E)$, $Z$ takes the form
\[
Z(\beta)=\int \dd E \; \rho(E) \; \e^{\beta E}\,.
\]
The vacuum expectation value (or ensemble average) of an observable
$O$ which is function of $E$ can be written as\footnote{The most
  general case in which $O(\phi)$ can not be written as a function of
  $E$ is discussed in Subsect.~\ref{subsect:observables}.}
\be 
\langle \mathcal{O}\rangle=\frac{1}{Z(\beta)}\int \dd E \; 
\mathcal{O}(E) \; \rho(E) \; \e^{\beta E}\,. 
\label{eq:exE}
\en 
Hence, a numerical determination of $\rho(E)$ would enable us to
express $Z$ and $\langle O \rangle$ as numerical integrals of known
functions in the single variable $E$. This approach is inherently different from
conventional Monte-Carlo calculations, which relie on the concept of
importance sampling, i.e. the configurations contributing to the
integral are generated with probability 
\[
P_\beta (E) \; =  \; \rho(E) \; \e^{\beta E}  / Z(\beta ) \,.
\]
Owing to this conceptual difference, the method we are proposing can
overcome notorious drawbacks of importance sampling techniques.

\subsection{The LLR method \label{sec:2.2} }
We will now detail our approach to the evaluation of the
density of states by means of a lattice simulations. 
Our initial assumption is that the density of states is a regular
function of the energy that can be always approximated in a finite
interval by a suitable functional expansion.
If we consider the energy interval $[E_k,E_k+\DE]$, under the physically
motivated assumption that the density of states is a smooth function
in this interval, the logarithm of the latter quantity can be written, using
Taylor's theorem, as  
\bea
\ln \, \rho (E) &=& \ln \, \rho \left(E_k +\frac{\DE}{2} \right)  \; +
\; \frac{  d \; \ln \, \rho }{dE } \Big\vert _{E=E_k+\DE/2} \;
\left(E- E_k -\frac{\DE}{2} \right) \; + \; R_k(E)  
\; ,   
\label{eq:kk1} \\ 
R_k(E) &=& \frac{1}{2} \, \frac{ d^2 \; \ln \, \rho }{dE^2 } \Big\vert
  _{E_k +\DE/2} \; \left(E- E_k -\frac{\DE}{2} \right) ^2 \; + \; {\cal O}(\DE^3)  \; . 
\nonumber  
\ena 
Thereby, for a given action $E$, the integer $k$ is chosen such that 
$$ 
E _k \leq E \leq E_k \, + \, \DE \; , \hbo 
E_k \; = \; E_0 \; + \; k \, \DE \; . 
$$
Our goal will be to devise a numerical method to calculate the 
Taylor coefficients 
\be 
a_k :=  \frac{ 
d \; \ln \, \rho }{dE } \Big\vert _{E=E_k+\DE/2} 
\label{eq:kk1a}
\en 
and to reconstruct from these an approximation for the density of states 
$\rho (E)$. 
By introducing the {\it intrinsic} thermodynamic quantities, $T_k$
(temperature) and $c_k$ (specific heat) by  
\be 
\frac{ d \; \ln \, \rho }{dE } \Big\vert _{E=E_k+\DE /2} \; = \; \frac{1}{T_k} \;
= \; a_k \;
, \hbo 
\frac{ d^2 \; \ln \, \rho }{dE^2 } \Big\vert _{E=E_k+\DE /2} \; = \; -
\frac{1}{T_k^2 \, c_k} \, \frac{1}{V} \; . 
\label{eq:kk2}
\en
we expose the important feature that the target coefficients $a_k$ are independent
of the volume while the correction  $R_k(E) $ is of order $\DE ^2/V$. 
In all practical applications, $R_k$ will be numerically
much smaller than $a_k \, \DE$. For a certain parameter range (i.e.,
for the correlation length smaller than the lattice size), we can
analytically derive this particular volume dependence of the density
derivatives. Details are left to the appendix. 

Using the trapezium rule for integration, we find in particular 
\bea 
\ln \, \frac{ \rho (E_{k+1} +\DE /2) }{ \rho(E_k +\DE /2)} &=& \int
_{E_k+\frac{\DE}{2}}^{E_{k+1}+\frac{\DE}{2}} \frac{ d \, \ln \rho }{dE} \; dE \; = \;
\frac{ \DE}{2} \, [ a_k   + a_{k+1} ] \; + \; {\cal O}(\DE^3) \; . 
\label{eq:kk3a}
\ena 
Using this equation recursively, we find 
\bea 
\ln \frac{ \rho (E_N +\frac{\DE}{2}) }{\rho (E_0 +\frac{\DE}{2})} &=&
\frac{a_0}{2} \,  \DE \; + \; \sum _{k=1}^{N-1} a_k \, \DE \; + \; \frac{a_N}{2} \, \DE \;  + \; 
 {\cal O}(\DE^2) \; . 
\label{eq:kk3b} 
\ena
Note that $N \, \DE = {\cal O}(1)$. Exponentiating
(\ref{eq:kk1}) and using (\ref{eq:kk3b}), we obtain   
\bea 
\rho (E) &=& \rho \left(E_N +\frac{\DE}{2}\right) \, \exp \Bigl\{ a_N  \, (E-E_N -\DE/2) \; + \; 
 {\cal O}(\DE^2) \Bigr\}  
\label{eq:kk4b} \\ 
&=&
\rho_0 \left( \prod_{k=1}^{N-1}  \e^{a_k \DE} \right) \;
\exp \left\{ a_N  \, \left(E  -E_N \right) \; + \; 
 {\cal O}(\DE^2) \right\}  \; , 
\label{eq:kk5}
\ena
where we have defined an overall multiplicative constant by
$$
\rho _0 \; = \; \rho \left(E_0 +\frac{\DE}{2}\right) \, \e ^{a_0
  \DE/2} \; . 
$$
We are now in the position to introduce the piecewise-linear and
continuous approximation of the density of states by 
\be
\tilde{\rho} (E) \; = \; \rho_0 \left( \prod_{k=1}^{N-1}  \e^{a_k \DE} \right) \; \e^{a_N (E-E_N) }  \; , \hbo N(E): \;
E_N \le E < E_{N+1} \; . 
\label{eq:rhoguess}
\en
i.e., $N$ is chosen in such a way that $  E_N \le
E < E_N +  \DE $ for a given $E$.
With this definition, we obtain the remarkable identity 
\be 
\rho (E) \; = \; \tilde{\rho } \left( E \right)  
\;  \exp \Bigl\{  {\cal O}(\DE^2) \Bigr\} \; = \; \tilde{\rho } \left( E \right)  
\Bigl[ 1 \; + \; {\cal O}(\DE^2) \Bigr] \; . 
\label{eq:kk4}
\en 
which we will extensively use below. We will observe that $\rho (E)$
spans many orders of magnitude. The 
key observation is that our approximation implements {\it exponential error
  suppression}, meaning that $\rho (E)$ can be approximated with nearly-constant
{\it relative error} despite it may reach over thousands of orders of
magnitude: 
\be 
1 - \frac{\tilde{\rho}(E)}{\rho (E)} \; = \; {\cal
  O}\left(  \DE^2  \right) \; . 
\label{eq:kk5b}
\en 

We will now present our method to calculate the coefficients $a_k$. 
To this aim, we introduce the action restricted and re-weighted
expectation values~\cite{Langfeld:2012ah} with $a$ being an external
variable: 
\bea
\dlangle W[\phi] \drangle_{k} (a)
&=& \frac{1}{{\cal N}_k} \int {\cal D} \phi \; \theta _{[E_k,\DE]}(S[\phi]) \; W[\phi] \; \, \; \; \e^{-a S[\phi] } \; ,
\label{eq:kk7} \\
{\cal N}_k &=& \int {\cal D} \phi \; \theta _{[E_k,\DE]} \; \, \; \; \e^{-a S[\phi] } \; = \; 
\int_{E_k}^{E_k+\DE}  \dd E \, \rho (E) \,\mathrm{e}^{-aE} \; , 
\label{eq:kk8} 
\ena
where we have used (\ref{eq:den}) to express $N_k$ as an ordinary
integral. We also introduced the modified Heaviside function 
$$ 
\theta _{[E_k,\DE]} (S) \; = \; \left\{ \begin{array}{ll} 
1 & \hbox{for} \; \; \; E_k \leq S \leq E_k + \DE \\ 
0 & \hbox{otherwise . } \end{array} \right. 
$$
If the observable only depends on the action, i.e., 
$ W[\phi] = O(S[\phi])$, (\ref{eq:kk7}) simplifies to 
\be 
\dlangle O \drangle_{k} (a)
\; = \; \frac{1}{{\cal N}_k} \int_{E_k}^{E_k+\DE}  \dd E \, \rho (E)
\; O(E) \; \,\mathrm{e}^{-aE} \; ,  
\label{eq:kk9} 
\en 
Let us now consider the specific action observable 
\bea
  \Delta E  = S - E_k - \frac{\DE}{2} \, , 
\label{eq:kk10}
\ena
and the solution $a$ of the non-linear equation 
\be 
\dlangle \Delta E \drangle_k (a) \; = \; 0 \; . 
\label{eq:kk11}
\en
Inserting $\rho (E)$ from (\ref{eq:kk4b}) into (\ref{eq:kk9}) and
defining $\Delta a = a_k - a$, we obtain: 
\bea 
\dlangle \Delta E \drangle_k (a) &=& \frac{ 
\rho (E_k+\DE /2) \,  \int _{E_k}^{E_k+\DE} 
dE \; (E- E_k - \DE/2) \; \e ^{ \Delta a \, (E-E_k) } \;
\e^{{\cal O}(\DE^2)} }{  \rho (E_k+\DE /2) \, \int _{E_k}^{E_k+\DE} 
dE \; \e ^{ \Delta a \,  (E-E_k) } \; \e^{{\cal O}(\DE^2)}  } 
\nonumber \\ 
&=& \frac{  \int _{E_k}^{E_k+\DE} 
dE \; (E- E_k - \DE/2) \; \e ^{ \Delta a \, (E-E_k) }  }{  
 \int _{E_k}^{E_k+\DE} dE \; \e ^{ \Delta a \,  (E-E_k) } \; 
 } \; + \; {\cal O}\Bigl(\DE^2\Bigr) \; =
\; 0 \; .
\label{eq:kk12} 
\ena 
Let us consider for the moment the function
$$
F(\Delta a) \; := \;  \frac{  \int _{E_k}^{E_k+\DE} 
dE \; (E- E_k - \DE/2) \; \e ^{ \Delta a \, (E-E_k) }  }{  
 \int _{E_k}^{E_k+\DE} dE \; \e ^{ \Delta a \,  (E-E_k) } \; 
} \; .
$$
It is easy to check that $F$ is monotonic and vanishing for $\Delta a
=0$: 
$$
F^\prime (\Delta a) \; > \; 0 \; , \hbo F(\Delta a=0) \; = \; 0 \; .
$$
We therefore conclude for any $\DE$ that if (\ref{eq:kk12}) does
have a solution, this solution is unique. For sufficiently small
$\DE$ there is a solution, and, hence, the only solution is given
by:  
\be 
\dlangle \Delta E \drangle_k (a) \; = \; 0 \hbo \Leftrightarrow \hbo 
a \; = \; \frac{ d \; \ln \, \rho }{dE } \Big\vert
_{E=E_k +\frac{\DE }{2}} \; + \; {\cal O}\Bigl(\DE^2\Bigr) \; .
\label{eq:kk15}
\en
The later equation is at the heart of the LLR algorithm: it details
how we can obtain the log-rho derivative by calculating the
Monte-Carlo  average $\dlangle \Delta E \drangle_k (a)$ (using
(\ref{eq:kk7})) and solving a non-linear equation, i.e.,
(\ref{eq:kk11}). 

\medskip 
In the following, we will discuss the practical implementation by
addressing two questions: (i) How do we solve the non-linear equation?
(ii) How do we deal with the statistical uncertainty since the
Monte-Carlo method only provides stochastic estimates for the
expectation value $\dlangle \Delta E \drangle_k (a)$? 

\medskip 
Let us start with the standard Newton-Raphson method to answer question
(i). Starting from an initial guess $a^{(0)}$ for the solution, this
method produces a sequence 
$$ 
a^{(0)} \; \to \; a^{(1)} \to \; a^{(2)} \to \; \ldots \; \to \;
a^{(n)} \to \; a^{(n+1)} \ldots \; , 
$$
which converges to the true solution $a_k$. Starting from $a^{(n)}$
for the solution, we would like 
to derive an equation that generates a value $a^{(n+1)}$ that is even
closer to the true solution: 
\be 
\dlangle \Delta E \drangle_k \Bigl(a^{(n+1)}\Bigr) \; = \; \dlangle
\Delta E \drangle_k \Bigl( a^{(n)} \Bigr) \; + \; 
\frac{d}{da} \dlangle \Delta E \drangle_k \Bigl( a^{(n)} \Bigr) \; 
\Bigl( a^{(n+1)} - a^{(n)} \Bigr) \; = \; 0 \; . 
\label{eq:kk16}
\en
Using the definition of $\dlangle \Delta E \drangle_k
\Bigl(a^{(n+1)}\Bigr) $ in (\ref{eq:kk12}) with reference to
(\ref{eq:kk10}) and (\ref{eq:kk9}), we find: 
\be 
\frac{d}{da} \dlangle \Delta E \drangle_k (a) \; = \; - \; \Bigl[ 
\dlangle \Delta E^2 \drangle_k (a) \; - \; \dlangle \Delta E
\drangle_k^2 (a) \, \Bigr] \; =: \; - \; \sigma ^2  (\Delta E; a) \; . 
\label{eq:kk17}
\en
We thus find for the improved solution: 
\be 
a^{(n+1)} \; = \; a^{(n)} \; + \;  \frac{ \dlangle \Delta E \drangle_k
  (a^{(n)})}{ \sigma ^2  (\Delta E; a^{(n)})}  \; . 
 \label{eq:kk18}
\en 
We can convert the Newton-Raphson recursion into a simpler fixed point
iteration if we assume that the choice $a^{(n)}$ is sufficiently close to the
true value $a_k$ such that  
$$ 
\DE \; \Bigl( a^{(n)} - a_k \Bigr) \; \ll \; 1 \; .  
$$
Without affecting the precision with which the solution $a$ of
(\ref{eq:kk12}) can be obtained, we replace 
\be 
\sigma ^2  (\Delta E; a) \; = \; \frac{1}{12}\,  \DE^2 \; \Bigl[ 1 \; +
\; {\cal O} \Bigl( \DE \Delta a \Bigr)^2 \Bigr] \; \; \Bigl[
1 + {\cal O}(\DE ) \Bigr] \; . 
\label{eq:kk19} 
\en 
Hence, the Newton-Raphson iteration is given by 
\be 
a^{(n+1)} \; = \; a^{(n)} \; + \;  \frac{12 }{\DE ^2 }  \; 
\dlangle \Delta E \drangle_k   (a^{(n)}) 
 \label{eq:kk20}
\en
We point out that one fixed point of the above iteration, i.e., 
$a^{(n+1)}=a^{(n)}=a$, is attained for  
$$ 
\dlangle \Delta E \drangle_k   (a) = 0 \; , 
$$ 
which, indeed, is the correct solution. We have already shown that the
above equation has only one solution. Hence, if the iteration
converges at all, it necessarily converges to the true solution. 
Note that convergence can always be achieved by suitable choice of 
under-relaxation. We here point out that the solution to question (ii)
above will involve a particular type of under-relaxation.

\medskip 
Let us address the question (ii) now. We have already pointed out that 
we have only a stochastic estimate for the expectation value $\dlangle
\Delta E \drangle_k   (a) $ and the convergence of the Newton-Raphson
method is necessarily hampered by the inevitable statistical error of
the estimator. This problem, however, has been already solved by
Robbins and Monroe~\cite{robbins1951}. 

For completeness, we shall now give a brief presentation of the algorithm.
The starting point is the function $M(x)$, and a constant $\alpha$, such that
the equation $M(x) = \alpha$ has a unique root at $x=\theta$. 
$M(x)$ is only available by stochastic estimation using the random
variable $N(x)$: 
$$ 
\mathbb E[N(x)] = M(x) \; , 
$$ 
with $\mathbb E[N(x)]$ being the ensemble average of $N(x)$.
The iterative root finding problem is of the type 
\be
x_{n+1} \; = \; x_n \; + \; c_n \, (\alpha-N(x_n))
\en
where $c_n$ is a sequence of positive numbers sizes satisfying the
requirements
\be
\sum^{\infty}_{n=0}c_n = \infty \quad \mbox{ and } \quad
\sum^{\infty}_{n=0}c^2_n < \infty
\en
It is possible to prove that under certain
assumptions~\cite{robbins1951}  on the function $M(x)$ 
the $\lim_{n\to\infty}x_n$ converges in $L^2$ and hence in probability
to the true value $\theta$.
A major advance in understanding the asymptotic properties of this
algorithm was the main result of~\cite{robbins1951}. If we restrict
ourselves to the case 
\be
c_n=\frac{c}{n}
\en
one can prove that $\sqrt{n}(x_n -\theta)$ is asymptotically normal 
with variance
\be
\sigma^2_x=\frac{c^2 \sigma^2_\xi}{2~c~M'(x)-1}
\label{eq:var}
\en
where $\sigma_\xi^2$ is the variance of the noise.
Hence, the optimal value of the constant $c$, which minimises the
variance is given by 
\be
c=\frac{1}{M'(\theta)} \; . 
\en
Adapting the Robbins-Monro approach to our root finding iteration in
(\ref{eq:kk20}), we finally obtain an under-relaxed Newton-Raphson
iteration 
\be 
a^{(n+1)} \; = \; a^{(n)} \; + \;  \frac{12 }{\DE ^2 \; (n+1) }  \; 
\dlangle \Delta E \drangle_k   (a^{(n)}) \; , 
 \label{eq:kk21}
\en
which is optimal with respect to the statistical noise during
iteration.

\subsection{Observables and convergence with $\DE$ \label{sec:2.3} }
\label{subsect:observables}

We have already pointed out that expectation values of observables
depending on the action only can be obtained by a simple integral over
the density of states (see~(\ref{eq:exE})). Here we develop
a prescription for determining the values of expectations of more
general observables by folding with the numerical density of states
and analyse the dependence of the estimate on $\DE$. 

\medskip 
Let us denote a generic observable by $B(\phi)$. Its expectation
value is defined by 
\be 
\langle B[\phi] \rangle \; = \; 
\frac{1}{Z(\beta)} \; \int {\cal D}\phi \; B[\phi] \;
\mathrm{e}^{\beta S[\phi ] }
\label{eq:l1}
\en 
In order to relate to the LLR approach, we break up the latter
integration into energy intervals: 
\be 
\langle B[\phi] \rangle \; = \; 
\frac{1}{Z(\beta)} \; \sum _i \int  {\cal
  D}\phi \; \theta _{[E_i, \DE ]} \; B[\phi] \; \;
\mathrm{e}^{\beta S[\phi ] } \; . 
\label{eq:l3}
\en
Note that $\langle B[\phi] \rangle$ {\it does not} depend on $\DE$. 

We can express $\langle B[\phi] \rangle$ in terms of a sum over
double-bracket expectation values by choosing 
$$
W := B[\phi] \;\exp \{ (\beta + a_i) S[\phi ] \} 
$$
in (\ref{eq:kk7}). Without any approximation, we find: 
\bea
\langle B[\phi] \rangle &=& \frac{1}{Z(\beta)} \; \sum _i 
{\cal N}_i \, \e^{a_iE_i} \;  \dlangle B[\phi] \; \exp \{ \beta S
[\phi ]\, + \, a_i ( S[\phi ] - E_i) \} \drangle \;  (E_i) , 
\label{eq:l5} \\ 
Z(\beta ) &=&  \sum _i {\cal N}_i \, \e^{a_iE_i} \;  \dlangle \; \exp
\{ \beta S [\phi ]\, + \, a_i ( S[\phi ] - E_i) \}  \drangle (E_i) \; .  
\label{eq:l6} 
\ena
where ${\cal N}_i = {\cal N}_i (a_i)$ is defined in (\ref{eq:kk8}). 
The above result can be further simplified by using (\ref{eq:kk4}): 
\bea 
{\cal N}_i \, \e^{a_iE_i} &=& \int _{E_i}^{E_i+\DE} dE\; \rho
(E) \; \exp \{-a_i (E-E_i)  \} \; = \; \e ^{ {\cal O}(\DE^2) }
\int _{E_i}^{E_i+\DE} dE\; \tilde{\rho} (E) \; \exp \{-a_i (E-E_i)  \} 
\nonumber \\
&=& \e ^{ {\cal O}(\DE^2) } \; \tilde{\rho} (E_i) \; 
\int _{E_i}^{E_i+\DE} dE\; = \; \DE \;  \tilde{\rho} \left(
E_i\right) \; \e^{ {\cal O}(\DE^2) }   
\nonumber \\ 
&=& \DE \;  \tilde{\rho} \left(E_i  \right) \; \Bigl[  1 
\; + \; {\cal O}( \DE ^2) \, \Bigr] \; . 
\label{eq:l7} 
\ena
We now define the approximation to $\langle B[\phi] \rangle $ by
\bea
\langle B[\phi] \rangle_\mathrm{app}  &=& \frac{1}{Z(\beta)} \, \sum _i 
\DE \;\tilde{\rho} \left(E_i\right) \;
\dlangle B[\phi] \, \exp \{ \beta S[\phi ]\, + \, a_i ( S[\phi ] -
E_i) \} \drangle   
\label{eq:l9} \\ 
Z(\beta ) &:=&  \sum _i\; \DE \; \tilde{\rho}\left(E_i \right) \;
\dlangle \; \exp \{ \beta S [\phi ]\, + \, a_i ( S[\phi ] 
- E_i) \}  \drangle .   
\label{eq:l10} 
\ena
Since the double-bracket expectation values do not produce a
singularity if $\DE  \to 0$, i.e., 
$$
\lim _{\DE \to 0} \, \dlangle B[\phi] \, \exp \{ \beta S
     [\phi ]\, +  \, a_i ( S[\phi ] - E_i) \} \drangle \; = \;
     \hbox{finite} \; , 
$$
using (\ref{eq:l7}), from (\ref{eq:l5}) and (\ref{eq:l6}) we find that 
\be
\langle B[\phi] \rangle \; = \; \langle B[\phi] \rangle_\mathrm{app}
\; + \; \sum _i {\cal O}(\DE^3) \; = \; 
\langle B[\phi] \rangle_\mathrm{app}
\; + \; {\cal O}(\DE^2) \; .
\label{eq:l10c} 
\en
The latter formula together with (\ref{eq:l9}) provides access to all
types of observables using 
the LLR method with little more computational resources: 
Once the Robbins-Monro iteration (\ref{eq:kk21}) has settled for an
estimate of the coefficient $a_k$, the Monte-Carlo simulation simply
continues to derive estimators for the double-bracket expectation
values in (\ref{eq:l9}) and (\ref{eq:l10}).

With the further assumption that the double-bracket expectation values 
are (semi-)positive, an even better error estimate is produced by our
approach: 
$$
\langle B[\phi] \rangle \; = \; \langle B[\phi] \rangle_\mathrm{app}
\; + \; \sum _i {\cal O}(\DE^3) \; = \; 
\langle B[\phi] \rangle_\mathrm{app} \; \Bigl[ 1 
\; + \; {\cal O}(\DE^2) \; \Bigr]  .
$$
This implies that the observable $\langle B[\phi] \rangle $ can be
calculated with an {\it relative error} of order $\DE ^2$. 
Indeed, we find  from (\ref{eq:l5},\ref{eq:l6},\ref{eq:l7}) that 
\bea
\langle B[\phi] \rangle &=& \frac{1}{Z(\beta)} \, \sum _i 
\DE \; \tilde{\rho} \left(E_i\right) \; 
\dlangle B[\phi] \, \exp \{ \beta S [\phi ]\, +
\, a_i ( S[\phi ] - E_i) \} \drangle  
\label{eq:l9b} \\ 
&\times & \exp \Bigl\{  {\cal O}(\DE^2) \,
\Bigr\} \; ,  
\nonumber \\ 
Z(\beta ) &:=&  \sum _i\; \DE \; \tilde{\rho} \left(E_i \right)
\; \dlangle  \; \exp \{ \beta S [\phi ]\, + 
\, a_i ( S[\phi ] - E_i) \}  \drangle  .  
\label{eq:l10b} 
\ena
Thereby, we have used 
\bea
\left\vert \sum _i a_i \, \exp \Bigl\{ c_i \DE^2 \Bigr\} \right\vert &
\leq & \sum _i \vert a_i \vert \, \left\vert \exp \{ c_i \DE^2 \} 
\right\vert   \; \leq \; \sum _i \vert a_i \vert \, \exp \{
c_\mathrm{max} \DE^2 \} 
\nonumber \\ 
&=& \exp \{ c_\mathrm{max} \DE^2 \} \; \sum _i a_i \; 
\; = \;  \exp \Bigl\{  {\cal O}(\DE^2) \, \Bigr\} \times \, \sum
_i a_i \; . 
\nonumber 
\ena 
The assumption of (semi-)positive double expectation values is true for
many action observables, and possibly also for Wilson loops, whose re-weighted and action restricted double
expectation values might turn out to be positive (as it is the case for their
standard expectation values). In this case, our method would
provide an efficient determination of those quantities. This is
important in particular for large Wilson loop expectation values,
since they are notoriously difficult to measure with importance sampling
methods (see e.g.~\cite{Luscher:2001up}). We also note that, in order
to have an accurate determination of a generic observable, any Monte-Carlo estimate of the
double expectation values must be obtained to good precision dictated
by the size of $\DE$. A detailed numerical investigation of these and
related issues is left to future work. 


\medskip 
For the specific case that the observable $B[\phi]$ only depends on the
action $S[\phi]$, we circumvent this problem and evaluate the
double-expectation values exactly. To this aim, we introduce for the
general case  $ \dlangle W[\phi] \drangle _k$ the generalised density
$w_k(E)$ by  
\be 
\rho (E) \; w_k (E) \; = \; \int {\cal D} \phi \; 
\theta _{[E_k,\DE]} (S[\phi]) \; W[\phi] \; \delta \Bigl( E -
S[\phi] \Bigr) \; . 
\label{eq:l15} 
\en
We then point out that if $W[\phi]$ is depending on the action only,
i.e., $W[\phi] = f(S[\phi])$, we obtain: 
$$ 
w_k(E) \; = \; f(E) \; \theta _{[E_k,\DE]} (E) \; . 
$$
With the definition of the double expectation value (\ref{eq:kk7}),
we find: 
\be 
 \dlangle W[\phi] \drangle _k (a_k) \; = \; \frac{ 
\int _{E_k}^{E_k+\DE} dE \; \rho (E) \; \e^{-a_k E} \; w_k(E) }{ 
\int _{E_k}^{E_k+\DE} dE \; \rho (E) \; \e^{-a_k E} }
\label{eq:l16}
\en 
Rather than calculating  $\dlangle W[\phi] \drangle _k$ by Monte-Carlo
methods, we can {\it analytically} evaluate this quantity (up to order ${\cal
  O}(\DE ^2) $ ). Using the observation that for any smooth ($C_2$)
function $g$  
$$ 
\int _{E_k}^{E_k+\DE} dE \; g(E) \; = \; \DE \; g \left( E_k + \frac{
    \DE}{2} \right) \; + \; {\cal O} \Bigl( \DE^3 \Bigr) \;
, 
$$
and using this equation for both, numerator and denominator of
(\ref{eq:l16}), we conclude  that 
\be 
\dlangle W[\phi] \drangle _k (a_k) \; = \; w_k \left( E_k + \frac{
    \DE}{2} \right) \; + \; {\cal O} \Bigl( \DE^2 \Bigr) \;
.
\label{eq:l17}
\en 
Let us now specialise to the case that is relevant for (\ref{eq:l9b})
with $B$ depending on the action only:
\bea 
W[\phi] &=& b\Bigl( S[\phi] \Bigr) \,\exp \{ \beta S [\phi ] +  a_i (
S[\phi ] - E_i) 
\} , 
\nonumber \\ 
w_i(E)&=& b(E) \, \exp \{ \beta E  +  a_i ( E - E_i) \}  .  
\label{eq:l18} 
\ena
This leaves us with
\be 
\dlangle W[\phi] \drangle _i (a_i) \; = \;
b\Bigl( E_i + \frac{\DE}{2} \Bigr) \; \e^{\beta ( E_i +
  \frac{\DE}{2} ) } \; \e^{a_i \frac{\DE}{2} } \; + \;
{\cal O} \Bigl( \DE^2 \Bigr) \; .
\label{eq:l18b} 
\en
Inserting (\ref{eq:l17}) together with (\ref{eq:l18}) into
(\ref{eq:l9}), we find: 
\bea
\langle B[\phi] \rangle &=& \frac{1}{Z(\beta)} \; \sum _i 
\DE \; \tilde{\rho} \left(E_i + \frac{\DE}{2} \right) \; b_i
\Bigl(E_i + \frac{\DE}{2} \Bigr) \; \e ^{\beta (E_i +
  \frac{\DE}{2})} \; + \; {\cal O}\Bigl( \DE^2 \Bigr) \; , 
\label{eq:l19} \\ 
Z(\beta) &=& \sum _i \DE \; \tilde{\rho} \left(E_i + \frac{\DE}{2} \right) \; \e ^{\beta (E_i  + \frac{\DE}{2})} \; . 
\label{eq:l20} 
\ena

\medskip 
Below, we will numerically test the quality of expectation values
obtained by the LLR approach using action
observables only, i.e., $B[\phi ] = O(S[\phi])$. We will find that we
indeed achieve the predicted precision in $\DE ^2 $ for this type of
observables (see below Fig.~\ref{fig:peakfigure}).

\subsection{The numerical algorithm}
So far, we have shown that a piecewise continuous approximation of the
density of 
states that is linear in intervals of sufficiently small amplitude
$\DE$ allows us to obtain a controlled estimate of averages of
observables and that the angular coefficients $a_i$ of the linear
approximations can be computed in each interval $i$ using the
Robbins-Monro recursion~(\ref{eq:kk21}). Imposing the continuity of
$\log \rho(E)$, one can then determine the latter quantity up to an
additive constant, which does not play any role in cases in which
observables are standard ensemble averages. 

The Robbins-Monro recursion can be easily implemented in a numerical
algorithm. Ideally,  the recurrence would be stopped when a tolerance
$\epsilon$ for $a_i$ is reached, i.e. when  
\be
\left| a^{(n+1)}_i- a^{(n)}_i \right| = \frac{12~\left| \Delta E_i(a^{(n)}_i) \right| }{(n+1)~\DE^2} \le \epsilon \ ,
\en
with (for instance) $\epsilon$ set to the precision of the
computation. When this condition is fulfilled, we can set  $a_i =
a^{(n+1)}_i$. However, one has to keep into account the fact that the
computation of $\Delta E_i$ requires an averaging over Monte-Carlo
configurations. This brings into play considerations about thermalisation
(which has to be taken into account each time we send $a^{(n)}_i \to
a^{(n+1)}_i$), the number of measurements used for determining $\Delta
E_i$ at fixed $a^{(n)}_i$ and -- last but not least -- fluctuations of
the $a^{(n)}_i$ themselves. 


Following those considerations, an algorithm based on the
Robbins-Monro recursion relation should depend on the following input
(tunable) parameters:
\begin{itemize}
\item $\NTHERM$, the number of Monte-Carlo updates in the restricted
  energy interval before starting to measure expectation values;
\item $\NSWEEPS$, the number of iterations used for computing
  expectation values;
\item $\NRM$, the number of Robbins-Monro iterations for determining
  $a_i$; 
\item $N_B$, number of final values from the  Robbins-Monro iteration
  subjected to a subsequent bootstrap analysis. 
\end{itemize}
The version of the LLR method proposed and implemented in this paper is reported in an
algorithmic fashion in the box Algorithm~\ref{algo:LLR}. This implementation
differs from that provided in~\cite{Langfeld:2012ah,Langfeld:2014nta}
by the replacement  of the originally proposed root-finding procedure based on a deterministic
Newton-Raphson like recursion with the Robbins-Monro recursion, which is better
suited to the problem of finding zeros of stochastic equations. 
\begin{algorithm}
\nonumber
  \KwIn{$\NSWEEPS$, $\NTHERM$, $\NRM$, $\NAVG$}
  \KwOut{$a_i$ $\forall i$}
    \For{$0 \le i < \left( \EMAX - \EMIN \right)/\DE$}
    {
      Initialise $E_i = \EMIN + i \DE$, $a_i^{0} = \bar{a}_i$;\\
      \For{$0 \le n <\NRM$}
      {
        \For{$k \le\NSWEEPS$}
        {
          Evolve the whole system with an importance sampling
          algorithm for one sweep according to the probability distribution
          \[
          P(E) \propto  e^{- a_i^{n} E}
          \]
          accepting only configuration such that $E_i \le E \le E_i 
            + \DE$\\
            \If{$j \ge \NTHERM$}
            {
              Compute $E^{(j)}$, the value of the energy in the current
              configuration $j$;
            }
          Compute 
          \[
          \Delta E_i(a_i^{(n)}) = \frac{1}{\NSWEEPS -
            \NTHERM} \left( \sum_{j> \NTHERM} E^{(j)} \right) - E_i - \frac{\DE}{2}
          \]
        }
        Compute
        \[
        a^{(n+1)}_i = a^{(n)}_i -  \frac{12 \Delta
            E_i(a^{(n)}_i) }{(n+1)~\DE^2} 
        \]
      }
      Repeat $N_B$ times to produce $N_B$ candidates $a_i$ for
      a subsequent bootstrap analysis
      }
\caption{The LLR method as implemented in this work.}
\label{algo:LLR}
\end{algorithm}
Since the $a_i$ are determined stochastically, a different reiteration
of the algorithm with different starting conditions and different
random seeds would produce a different value for the same $a_i$.  The
stochastic nature of the process implies that the distribution of
the $a_i$ found in different runs is Gaussian. The generated ensemble
of the $a_i$ can then be used to determine the error of the estimate of
observables using analysis techniques such as jackknife and bootstrap. 

The parameters $\EMIN$ and $\EMAX$ depend on the system and on the
phenomenon under investigation. In particular, standard thermodynamic
considerations on the infinite volume limit imply that if one is interested in
a specific range of temperatures and the studied observables
can be written as statistical averages with Gaussian fluctuations, it is possible to restrict the
range of energies between the energy that is typical of the smallest
considered temperature and the energy that is typical of the highest
considered temperature. Determining a reasonable value for the
amplitude of the energy interval $\DE$ and the other tunable
parameters $\NSWEEPS$, $\NTHERM$, $\NRM$ and $\NAVG$ requires a modest
amount of experimenting with trial values. In our applications we
found that the results were very stable  for wide ranges of values of
those parameters. Likewise, $\bar{a}_i$, the initial value for the Robbins-Monro
recursion in interval $i$, does not play a crucial role; when
required and possible, an initial value close to the expected result
can be inferred inverting $\langle E (\beta)\rangle $, which
can be obtained with a quick study using conventional techniques.

The average $\dlangle \dots \drangle$ imposes an
update that restricts configurations to 
those with energies in a specific range. In most of our
studies, we have imposed the constraint analytically at the
level of the generation of the newly proposed variables, which results
in a performance that is comparable with that of the unconstrained
system. Using a simple-minded more direct approach, in which one
imposes the constraint after the generation of the proposed new variable, we
found that in most cases the efficiency of Monte-Carlo
algorithms did not drop drastically as a consequence of the
restriction, and even for systems  like SU(3) (see
Ref.~\cite{Langfeld:2012ah}) we 
were able to keep an efficiency of at least 30\% and in most cases no
less than 50\% with respect to the unconstrained system.  

\subsection{Ergodicity \label{sec:2.5} }

\begin{figure}[t]
\begin{center}
\includegraphics[width=0.4\textwidth]{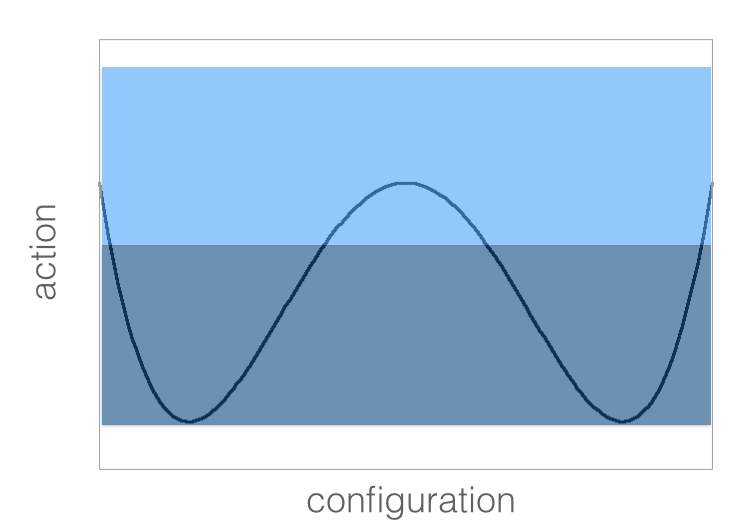} 
\includegraphics[width=0.4\textwidth]{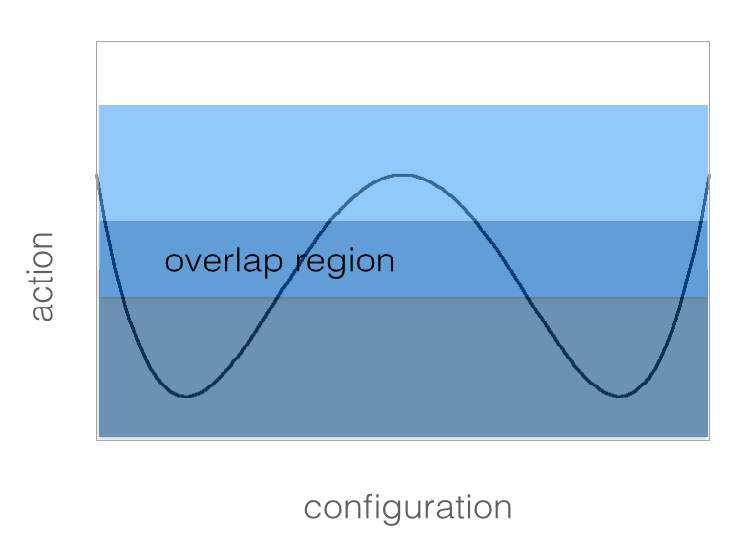} 
\end{center}
\caption{    Left: For contiguous energy intervals if a
    transition between configurations with energy in the same interval 
    requires going through configurations with energy that are outside that
    interval, the simulation might get trapped in one of the allowed
    regions (in green). Right: For overlapping energy intervals with replica
    exchange, the simulation can travel from one allowed region to the
    other through excursions to the upper interval.} 
\label{fig:ergodicity}
\end{figure}
Our implementation of the energy restricted average $\dlangle\cdots
\drangle$ assumes that the update algorithm is able to generate all
configurations with energy in the relevant interval starting from
configurations that have energy in the same interval. This assumption
might be too strong when the update is local\footnote{This is for
  instance the case for the popular heath-bath 
  and Metropolis update schemes.} in the energy (i.e. each elementary
update step changes the energy by a quantity of order
one for a system with total energy of order
$V$) and  there are topological excitations that can create
regions with the same energy that are separated by high energy
barriers. In these cases, which are rather common in gauge theories and
statistical mechanics\footnote{For instance, in a $d$-dimensional
  Ising system of size $L^d$, to  go from one groundstate to the other
  one needs to create a kink, which has energy growing as $L^{d-1}$.},
generally in order to go from one acceptable region to the other one
has to travel through a region of energies that is forbidden by an
energy-restricted update method such as the LLR. Hence, by construction,
in such a scenario our algorithm will get trapped in one of the
allowed regions. Therefore, the update will not be ergodic. 

In order to solve this problem, one can use an adaptation of
the replica exchange method~\cite{Replica}, as first proposed
in~\cite{Replicawl}.  The idea is that instead of dividing the whole
energy interval in contiguous sub-intervals overlapping only in one
point (in the following simply referred to as contiguous intervals),
one can divide it in 
sub-intervals overlapping in a finite energy region (this case will be
referred to as overlapping intervals). With the latter prescription, after a
fixed number of iterations of the Robbins-Monro procedure, we can check
whether in any pairs of 
overlapping intervals $(I_1, I_2$) the energy of both the corresponding
configurations is in the common region. For pairs fulfilling this
condition, we can propose an exchange of the configurations with a
Metropolis probability
\be
P_{\mathrm{swap}} = \mathrm{min}\left(1,e^{\left(a^{(n)}_{I_1} -
      a^{(n)}_{I_2}\right) 
  \left(E_{C_1} - E_{C_2}\right)}\right)  \ ,
\en
where $a^{(n)}_{I_1}$ and $a^{(n)}_{I_2}$ are the values
of the parameter $a$ at the current $n$-th iterations of the
Robbins-Monro procedure respectively in intervals $I_1$ and $I_2$ and
$E_{C_1}$ ($E_{C_2}$)  is the value of the energy of the
current configuration $C_1$ ($C_2$) of the replica in the interval
$I_1$ ($I_2$). If the proposed exchange is accepted, $C_1 \to C_2$ and
$C_2 \to C_1$. With repeated exchanges of configurations from
neighbour intervals, the system can now travel through all
configuration space. A schematic illustration of how this mechanism works is
provided in Fig.~\ref{fig:ergodicity}.   

As already noticed in~\cite{Replicawl}, the replica exchange step is
amenable to parallelisation and hence can be conveniently deployed in
calculations on massively parallel computers. Note that the replica
exchange step adds another tunable parameter to the algorithm, which
is the number $N_{\mathrm{SWAP}}$ of configurations swaps during the
Monte-Carlo simulation at a given Monte-Carlo step. A modification of
the LLR algorithm that incorporates this step can be easily implemented.

\subsection{Reweighting with the numerical density of
  states \label{sec:2.6}}  

In order to screen our approach outlined in subsections \ref{sec:2.2}
and \ref{sec:2.3} for ergodicity violations and to propose an efficient
procedure to calculate any observable once an estimate for the density
of states has been obtained, as an alternative to the replica
exchange method discussed in the previous section, we here introduce
an importance sampling algorithm with reweighting with respect to the
estimate $\tilde{\rho}$. This algorithm features short correlation times
even near critical points. Consider for instance a system described by
the canonical ensemble. We define a modified Boltzmann weight $W_B(E)$
as follows:  
\be
W_B(E) = 
\left\{
\begin{array}{ll}
e^{ - \beta_1 E + c_1} & \mbox{for} \ E < \EMIN \ ; \\
1/\tilde{\rho}(E) & \mbox{for} \ \EMIN \le E \le \EMAX \ ; \\
e^{ - \beta_2 E + c_2} & \mbox{for} \ E > \EMAX \ .
\end{array}
\right.
\label{eq:4.1}
\en
Here $\EMIN$ and $\EMAX$ are two values of the energy that are far from
the typical energy of interest $E$: 
\be
\EMIN \ll E \ll \EMAX \ .
\en
If conventional Monte-Carlo simulations can be used for numerical
studies of the given system, we can chose $\beta_1$ and $\beta_2$ from
the conditions  
\be
\langle E(\beta_i) \rangle = E_i \ , \qquad i=1,2 \ .
\en
If importance sampling methods are inefficient or unreliable,
$\beta_1$ and $\beta_2$ can be chosen to be the micro-canonical
$\beta_{\mu}$ corresponding respectively to the density of states
centred in $\EMIN$ and $\EMAX$. These $\beta_{\mu}$ are outputs of our
numerical determination $\tilde{\rho}(E)$. The two constants $c_1$ and
$c_2$ are determined by requiring continuity of $W_B(E)$ at $\EMIN$ and
at $\EMAX$: 
\be
\lim_{E \to \EMIN^-} W_B(E)= \lim_{E \to \EMIN^+} W_B(E)\qquad \mbox{and}
\qquad \lim_{E \to \EMAX^-} W_B(E)= \lim_{E \to \EMAX^+}  W_B(E)  \ . 
\en 
Let $\rho(E)$ be the correct density of state of the
system. If $\tilde{\rho}(E) = \rho(E)$, then for $\EMIN \le E \le
\EMAX$ 
\be
\rho(E) W_B(E) = 1 \ ,
\en
and a Monte-Carlo update with weights $W_B(E)$ drives the system in
configuration space following a random walk in the energy. In
practice, since $\tilde{\rho}(E)$ is determined numerically, upon
normalisation 
\be
\rho(E) W_B(E) \simeq 1 \ ,
\en
and the random walk is only approximate. However, if $\tilde{\rho}(E)$ is a
good approximation of $\rho(E)$, possible free energy
barriers and metastabilities of the canonical system can be
successfully overcome with the weights~(\ref{eq:4.1}). Values of
observables for the canonical ensemble at temperature $T = 1/\beta$
can be obtained using reweighting: 
\be
\langle O (\beta) \rangle = \frac{\langle O e^{- \beta E}
  (W_B(E))^{-1} \rangle_W}{\langle e^{- \beta E} (W_B(E))^{-1}
  \rangle_W} \ , 
\label{eq:4.7}
\en
where $\langle \ \rangle$ denotes average over the canonical ensemble
and $\langle \ \rangle_W$ average over the modified ensemble defined
in~(\ref{eq:4.1}). The weights $W_B(E)$ guarantees ergodic sampling
with small auto-correlation time for the configurations with energies
$E$ such that $\EMIN \le E \le \EMAX$, while suppressing to energy $E \ll
\EMIN$ and $E \gg \EMAX$. Hence, as long as for a given $\beta$ of the
canonical system $\overline{E} = \langle E \rangle$ and the energy
fluctuation $\langle \Delta E = \sqrt{\langle (E - \langle E \rangle)^2
  \rangle}$ are such that  
\be
\EMIN \ll \langle E \rangle - \Delta \overline{E} \qquad \mbox{and} \qquad
\langle E \rangle + \Delta \overline{E} \ll \EMAX \ ,  
\label{eq:4.8}
\en
the reweighting~(\ref{eq:4.7}) does not present any overlap
problem. The role of $\EMIN$ and $\EMAX$ is to restrict the approximate
random walk only to energies that are physically interesting, in order
to save computer time. Hence, the choice of $\EMIN$, $\EMAX$ and of the
corresponding $\beta_1$, $\beta_2$ do not need to be fine-tuned, the
only requirement being that Eqs.~(\ref{eq:4.8}) hold. These conditions
can be verified {\em a posteriori}. Obviously, choosing the smallest
interval $\EMAX - \EMIN$ where the conditions~(\ref{eq:4.8}) hold
optimises the computational time required by the algorithm. The
weights~(\ref{eq:4.7}) can be easily imposed using a metropolis or a
biased metropolis~\cite{Bazavov:2005zy}. Again, due to the absence of
free energy barriers, no ergodicity problems are expected to
arise. This can be checked by verifying that in the simulation there
are various {\em tunnellings} (i.e. round trips) between $\EMIN$ and
$\EMAX$ and that the frequency histogram of the energy is approximately
flat between $\EMIN$ and $\EMAX$. Reasonable requirements are to have
${\cal O}(100-1000)$ tunnellings and an histogram that is flat within
15-20\%. These criteria can be used to confirm that the numerically
determined $\rho(E)$ is a good approximation of
$\overline{\rho}(E)$. The flatness of the histogram is not influenced
by the $\beta$ of interest in the original multi-canonical
simulation. This is particularly important for first order phase
transitions, where traditional Monte-Carlo algorithms have a tunnelling
time that is exponentially suppressed with the volume of the
system. Since the modified ensemble relies on a random walk in energy,
the tunnelling time between two fixed energy densities is expected to
grow only as the square root of the volume.\\ 
This procedure of using a modified ensemble followed by
reweighting is inspired by the multi-canonical
method~\cite{Berg:1992qua}, the only substantial difference being the
recursion relation for determining the weights. Indeed for U(1)
lattice gauge theory a multi-canonical update for which the weights are
determined starting from a Wang-Landau recursion is discussed
in~\cite{Berg:2006hh}. We also note that the procedure used here
to restrict ergodically the energy interval between $\EMIN$
and $\EMAX$ can be easily implemented also in the replica exchange
method analysed in the previous subsection.

\section{Application to Compact U(1) Lattice Gauge Theory} 
\label{sect:u1}

\subsection{The model}
Compact U(1) Lattice Gauge Theory is the simplest gauge theory based
on a Lie group. Its action is given by 
\bea
S = \beta \sum_{x, \mu < \nu}  \cos(\theta_{\mu \nu}(x) ) \ ,
\ena
where $\beta = 1/g^2$, with $g^2$ the gauge coupling, $x$ is a point
of a d-dimensional lattice of size $L^d$ and  $\mu$ and $\nu$ indicate
two lattice directions, indicised from $1$ to $d$ (for simplicity, in
this work we shall consider only the case $d = 4$), $\theta_{\mu \nu}$
plays the role of the electromagnetic field tensor: if we associate
the compact angular variable $\theta_{\mu}(x) \in [ - \pi; \pi [$ with
the link stemming from $i$ in direction $\hat{\mu}$,  
\beq
\theta_{\mu \nu}(x) = \theta_{\mu}(x) + \theta_{\nu}(x + \hat{\mu}) -
\theta_{\mu}(x + \hat{\nu}) - \theta_{\nu}(x) \ . 
\eeq
The path integral of the theory is given by
\beq
Z = \int {\cal D \theta_{\mu}}\;  \e^{S} \, , \hbo 
{\cal D \theta_{\mu}} = \prod_{x,\mu} \frac{\mbox{d}
  \theta_{\mu}(x)}{2 \pi} \ ,
\eeq
the latter identity defining the Haar measure of the U(1) group.\\
The connection with the framework of SU(N) lattice gauge theories is
better elucidated if we introduce the link variable 
\beq
U_{\mu}(x) = e^{i \theta_{\mu}(x)} \ .
\eeq
With this definition, $S$ can be rewritten as
\beq
S = \beta \sum_{x, \mu < \nu}  {\cal R}e \, P_{\mu \nu} (x)\ ,
\eeq
with 
$$
P_{\mu \nu}(x) = U_{\mu}(x) U_{\nu}(x + \hat{\mu}) U_{\mu}^{\ast}
(x + \hat{\nu}) U_{\nu}^\ast (x)
$$ 
the plaquette variable and $U^{\star}_{\mu}(x)$ is the complex
conjugate of $U_{\mu}(x)$. Working with the variables $U_{\mu}(x)$
allows us to show immediately that $S$ is invariant under U(1) gauge
transformations, which act as  
\beq
U_{\mu}(x) \mapsto \Lambda^{\ast}(x) \; U_{\mu}(x) \; \Lambda(x + \hat{\mu})
\ , \qquad \Lambda(x) = e^{i \lambda(x)} \ , 
\eeq 
with $\lambda(x) \in [-\pi; \ \pi[$ a function defined on lattice points.  

The connection with U(1) gauge theory in the continuum can be shown by
introducing the lattice spacing $a$ and the non-compact gauge field
$a \, A_{\mu}(x) = \theta_{\mu}(x)/ g$, so that 
\beq
U_{\mu}(x) = e^{i g a \, A_{\mu}(x)}  \ .
\eeq
Taking $a$ small and expanding the cosine leads us to
\beq
S = - \frac{1}{4} a^4 \sum_{x, \mu, \nu} \left( \Delta_{\mu} A_{\nu}(x)
  - \Delta_{\nu} A_{\mu}(x) \right)^2 + {\cal O}(a^6) \, + \,
\hbox{constant} \; , 
\eeq
with $\Delta_{\mu}$ the forward  difference operator. 
In the limit $a \to 0$, we finally find 
\beq
S \simeq - \frac{1}{4} \int \mbox{d}^4 x F_{\mu \nu}(x) ^2 \, , 
\eeq
with $F_{\mu \nu}$ being the usual field strength tensor. 
This shows that in the {\em classical} $a \to 0$ limit $S$ becomes the
Euclidean action of a free gas of photons, with interactions being
related to the neglected lattice corrections. It is worth to remark
that this classical continuum limit is not the continuum limit of the
full theory. In fact, this classical continuum limit is spoiled by
quantum fluctuations. These prevent the system from developing a
second order transition point in the $a \to 0$ limit, which is a
necessary condition to be able to remove the ultraviolet cutoff
introduced with the lattice discretisation. The lack of a continuum
limit is related to the fact that the theory is strongly coupled in
the ultraviolet. Despite the non-existence of a continuum limit for
Compact U(1) Lattice Gauge Theory, this lattice model is still interesting,
since it provides a simple realisation of a weakly first order phase
transition. This bulk  phase transition separates a confining phase at
low $\beta$ (whose existence was pointed out by
Wilson~\cite{Wilson:1974sk} in his seminal work on Lattice Gauge
Theory) from a deconfined phase at high $\beta$, with the transition itself occurring at  a critical
value of the coupling $\beta_c \simeq 1$.  Rather unexpectedly at
first side, importance-sampling Monte-Carlo studies of this phase
transitions turned out to be demanding and not immediate to interpret,
with the order of the transition having been debated for a long time
(see
e.g.~\cite{Creutz:1979zg,Lautrup:1980xr,Bhanot:1981zg,Jersak:1983yz,Azcoiti:1991ng,Bhanot:1992nf,Lang:1994ri,Kerler:1995va,Jersak:1996mn,Campos:1998jp}). The 
issue was cleared only relatively recently, with investigations that
made a crucial use of
supercomputers~\cite{Arnold:2000hf,Arnold:2002jk}. What makes the
transition difficult to observe numerically is the role played in the
deconfinement phase transition by magnetic
monopoles~\cite{frolich:1987}, which condense in the confined
phase~\cite{frolich:1987,DiGiacomo:1997sm}. 

The existence of topological sectors and the presence of a transition
with exponentially suppressed tunnelling times can provide robust
tests for the efficiency and the ergodicity of our algorithm. This
motivates our choice of Compact U(1) for the numerical investigation
presented in this paper.   

\subsection{Simulation details}

\begin{figure}[t]
\begin{minipage}[l]{7.5cm}
 \begin{center}
\includegraphics[width=8.5cm]{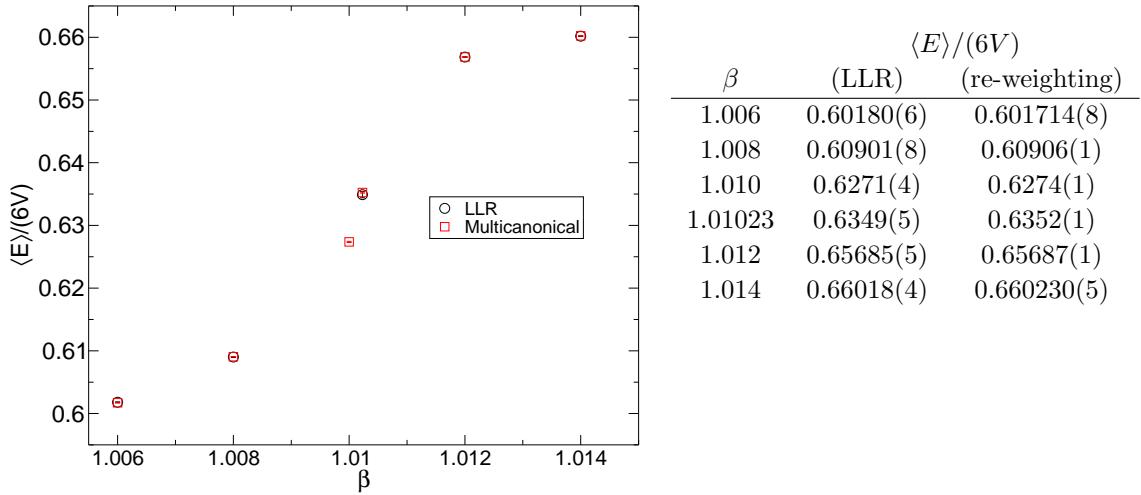} 
\end{center}
\end{minipage}
\hspace{1cm} 
\begin{minipage}[r]{4cm}
\protect\small
\begin{tabular}{c c c}
  &  \multicolumn{2}{c}{$\langle E \rangle/(6V) $ }    \\ 
$ \beta $ &   (LLR)  &   (re-weighting)   \\ 
  \hline
  1.006   & 0.60180(6) & 0.601714(8)  \\
  1.008   & 0.60901(8) & 0.60906(1)   \\
  1.010   & 0.6271(4)  & 0.6274(1) \\
  1.01023 & 0.6349(5)  & 0.6352(1) \\
  1.012   &  0.65685(5)& 0.65687(1) \\
  1.014   &  0.66018(4)& 0.660230(5) \\
\end{tabular} 
\vspace{2cm}
\end{minipage}
\caption{Comparison between the plaquette computed with the LLR
  algorithm (see subsection~\ref{sec:2.2}) and via re-weighting with
  respect to the estimate $\tilde{\rho}$ (see
  subsection~\ref{sec:2.6}) for a $L=12$ lattice. 
\label{fig:plaq_comp} }
\end{figure}

The study of the critical properties of U(1) lattice gauge theory is
presented in this section. In order to test our algorithm, we
investigated the behaviour of specific heath as function of the volume.
This quantity has been carefully investigated in previous studies, and
as such provides a stringent test of our procedure. In order to
compare data across different sizes, our results will be often
provided normalised to the number of plaquette $6 L^4 = 6V$.

\begin{figure}[t]
\begin{center}
\includegraphics[width=0.9\textwidth]{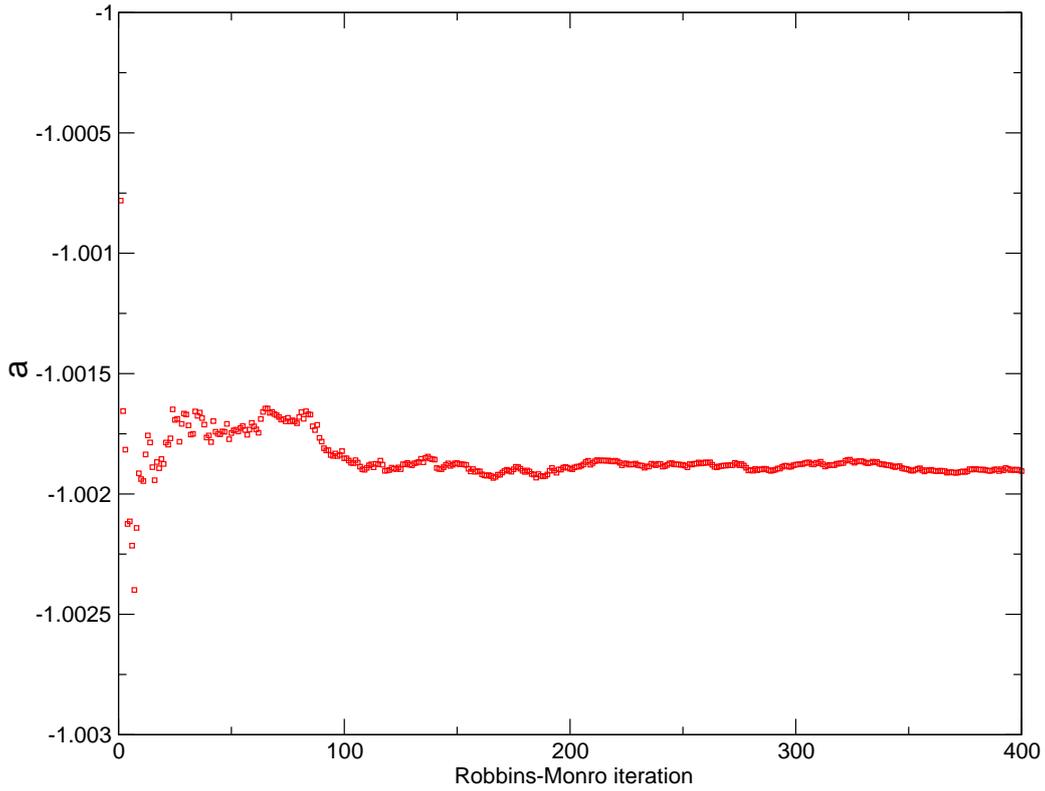} 
\end{center}
\caption{Estimated $a_i$ as a function of the Robbins-Monro
  iteration, on a $20^4$ lattice and for action $E/(6V) = 0.59009548$ at the centre of
  the interval with $\DE/V=1.91\times 10^{-4}$.}
\label{fig:history_rm}
\end{figure}

We studied lattices sizes ranging from $8^4$ to $20^4$ and for each
lattice size we computed the density of states $\rho(E)$ in the entire
interval $E_{\mbox{min}}\le E \le E_{\mbox{max}}$ (see
Tab.~\ref{tab:SimDet}). The rational behind the choice of the energy
region is that it must be centred  around the critical energy and it
has to be large enough to study all the critical properties of the
theory, i.e. every observable evaluated has to have support in this
region and have virtually no correction coming from the choice of the
energy boundaries.  
\begin{table}[t]
\begin{center}
\begin{tabular}{c c c c c c}
  L & $E_{\rm min}/(6V)$ & $E_{\rm max}/(6V)$ & $N_{SW}$ 
   &  $N_{RM}$ &$(E_{\rm max}-E_{\rm min})/\DE$\\
  \hline
  8  & 0.5722222 & 0.67  &250 & 600 & 512\\
  10,12,14,16,18,20 & 0.59& 0.687777 & 200 & 400 & 512  \\
  \end{tabular}
\caption{Values of the tunable parameters of the LLR algorithm used in our numerical
investigation.}
\label{tab:SimDet}
\end{center}
 \end{table}
We divided the energy interval in steps of $\DE$ 
and for each of the sub-interval we have repeated the entire generation of the
log-linear density of states function and evaluation of the observables
$N_{B}=20$ times to create the bootstrap samples
for the estimate of the errors. The values of the other tunable
parameters of the algorithm used in our study are reported in
Tab.~\ref{tab:SimDet}. An example determination of one of the $a_i$ is
reported in Fig.~\ref{fig:history_rm}. The plot shows the rapid
convergence to the asymptotic value and the negligible amplitude of
residual fluctuations. Concerning the cost of the simulations, we found that accurate
determinations of observables can be obtained with modest computational
resources compared to those needed in investigations of the system
with importance sampling methods. For instance, the most costly
simulation presented here, the investigation of the $20^4$ lattice,
was performed on 512 cores of Intel Westmere processors in about five
days. This needs to be contrasted with the fact that in the early
2000's only lattices up to $18^4$ could be reliably investigated with
importance sampling methods, with the largest sizes requiring
supercomputers~\cite{Arnold:2000hf,Arnold:2002jk}. 

One of our first analyses was a screening for potential ergodicity violations with
the LLR approach. As detailed in subsection~\ref{sec:2.5}, these can
emerge for LLR simulations using  contiguous intervals as it is the
case for the U(1) study reported in this paper. To this aim, we
calculated the action expectation value $\langle E \rangle$ for a
$12^4$ lattice for several values using the LLR method and using the
re-weighting with respect to the estimate $\tilde{\rho}$. Since the
latter approach is conceptually free of ergodicity issues, any
violations by the LLR method would be flagged by discrepancy. Our
findings are summarised in Fig.~\ref{fig:plaq_comp} and the
corresponding table. We find good agreement for the results from both
methods. This suggests that topological objects do not generate energy
barriers that trap our algorithm in a restricted section of
configuration space. Said in other words, for this system the LLR
method using contiguous interval seems to be ergodic.

\subsection{Volume dependence of $\log \tilde{\rho}$ and computational
  cost of the algorithm}
\begin{figure}[t]
\begin{center}
\includegraphics[height=5.7cm]{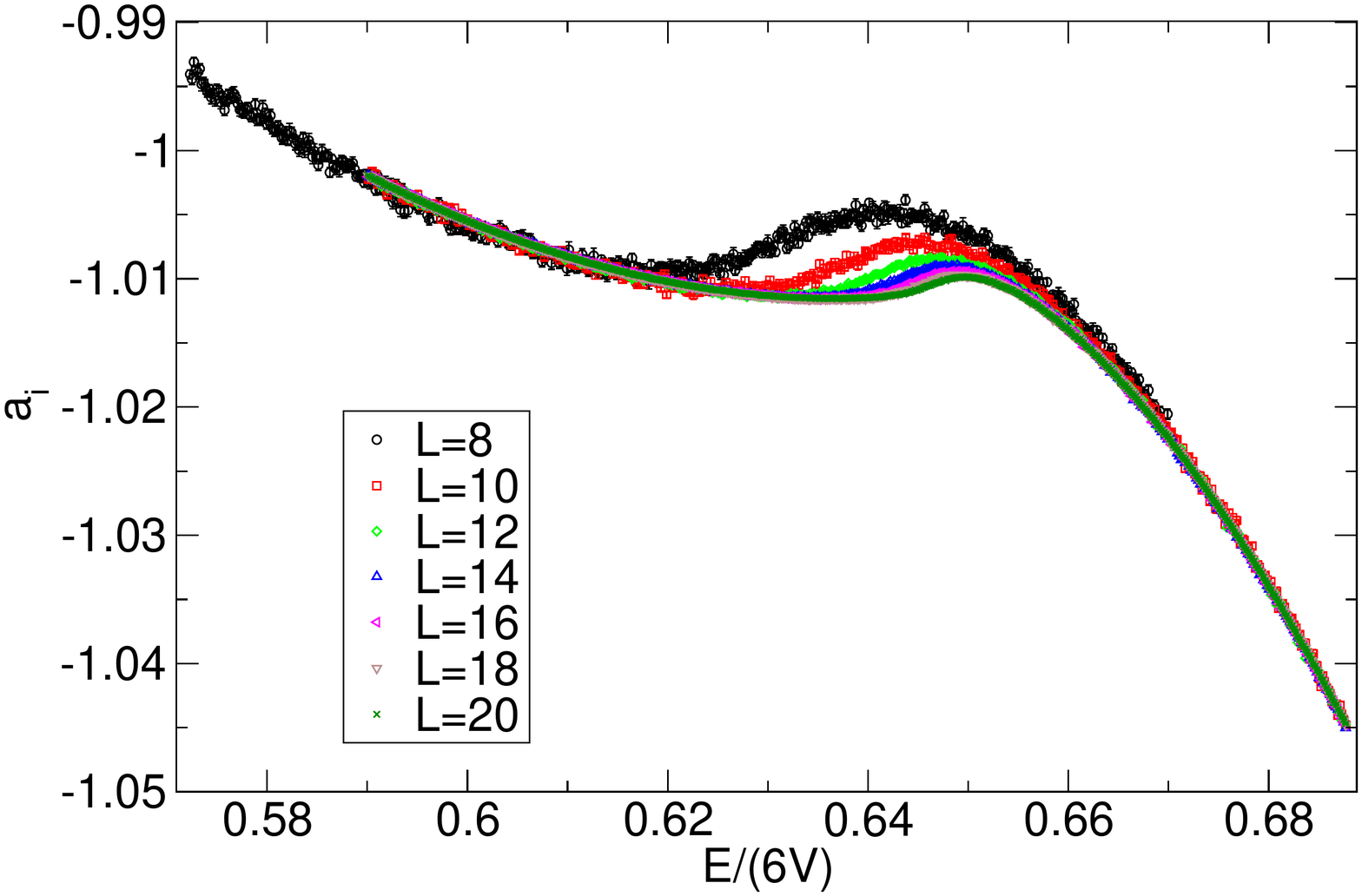} 
\includegraphics[height=5.7cm]{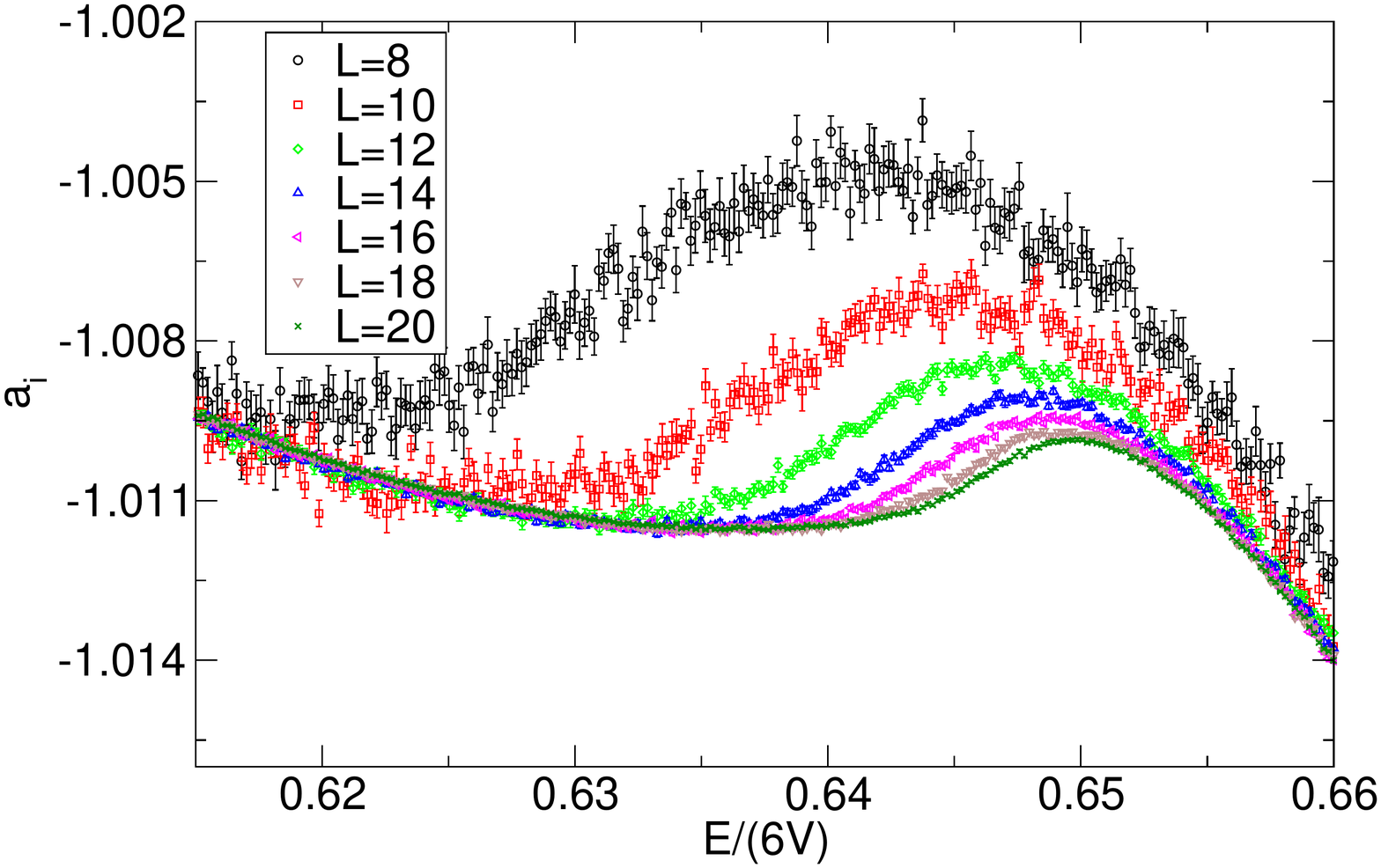} 
\end{center}
\caption{Estimate of $a_i$ as function of the energy density for various volume,
  the right panel is a zoom of the interesting region.}
\label{fig:ai}
\end{figure}

As first investigation we have performed a study of the scaling
properties of the $a_i$ as function of the volume. In Fig.~\ref{fig:ai}
we show the behaviour of the $a_i$ with the lattice volume. The
estimates are done for a fixed $\DE/V$, where the chosen value for the
ratio fulfils the request that within the errors all our 
observables are not varying for $\DE\to 0$ (we report on the study of
$\DE\to 0$ in section \ref{sec:descaling}). As it is clearly visible
from the plot the data are scaling toward a infinite volume estimate
of the $a_i$ for fixed energy density. 

As mentioned before, the issue facing importance sampling studies at
first order phase transitions are connected with tunnelling times that grow exponentially
with the volume. With the LLR method, the algorithmic cost is expected
to grow with the size of the system as $V^2$, where one factor of $V$
comes from the increase of the size and the other factor of $V$
comes from the fact that one needs to keep the energy interval per
unit of volume $\DE/V$ fixed, as in the large-volume limit only
intensive quantities are expected to determine the physics. One might
wonder whether this apparently simplistic argument fails at the first
order phase transition point. This might happen if the dynamics is such
that a slowing down takes place at criticality. In the case of Compact
U(1), for the range of lattice sizes studied here, we have found that
the computational cost  of the algorithm is compatible with a
quadratic increase with the volume.

\subsection{Numerical investigation of the phase transition}
Using the density of states it is straightforward to evaluate, by
direct integration (see subsection \ref{sec:2.3}), the
expectation values of any power of the energy and evaluate
thermodynamical quantities like the specific heat
\be
C_V(\beta) = \langle E^2(\beta)\rangle-\langle E(\beta)\rangle^2
\en
As usual we define the pseudo-critical coupling $\beta_c(L)$ such as
the coupling at which the peak of the specific heat occurs for a fixed
volume. The peak of the specific heat has been located using our
numerical procedure and the error bars are computed using the bootstrap
method.  Our results are summarised in Tab.~\ref{tab:peakloc} with a
comparison with the values in \cite{Arnold:2000hf}. Once again, the
agreement testify the good ergodic properties of the algorithm.
\begin{table}[hb]
\begin{center}
\begin{tabular}{c  c  c }
  L & $ \beta_{c}(L)$ present method & $\beta_{c}(L)$ reference values \\
  \hline
  8  & 1.00744(2) & 1.00741(1)  \\
  10 & 1.00939(2) & 1.00938(2)\\
  12 & 1.010245(1) & 1.01023(1)\\
  14 & 1.010635(5)& 1.01063(1)\\
  16 & 1.010833(4)& 1.01084(1)\\
  18 & 1.010948(2)& 1.010943(8)\\
  20 & 1.011006(2)&  \\
  \end{tabular} 
  \caption{$\beta_c(L)$ evaluated with the LLR algorithm
    and reference data from \cite{Arnold:2000hf}.}
\label{tab:peakloc}
\end{center}
\end{table}
Using our data it is possible to make a precise estimate of the
infinite volume critical beta by means of a finite size scaling analysis.
 The finite size scaling of the pseudo-critical coupling is given by
 \be
 \beta_{c}(L)=\beta_{c}+\sum_{k=1}^{k_{max}} B_k L^{-4 k} 
\label{eq:couplingscaling},
 \en
 where $ \beta_{c} $ is the critical coupling.
 We fit our data with the function in Eq.~(\ref{eq:couplingscaling}), the results are reported in Tab.~\ref{tab:betac}.

\begin{table}
\begin{center}
\begin{tabular}{c  c  c c }
  $ L_{min} $ & $ k_{max} $ &  $ \beta_{c} $ & $\chi^{2}_{red}$  \\
  \hline
  \textbf{14}  & \textbf{1} & \textbf{1.011125(3)} & \textbf{0.91}  \\
  \hline
  12 & 1 & 1.011121(3) & 2.42\\
  \textbf{12} & \textbf{2} & \textbf{1.011129(4)} & \textbf{0.67}\\
  \hline
  10 & 1 & 1.011116(5)& 7.44\\
  \textbf{10} & \textbf{2} & \textbf{1.011127(3)} & \textbf{0.60}\\
  \hline
  8 & 1 & 1.011093(5)& 90.26\\
  \textbf{8} & \textbf{2} & \textbf{1.011126(2)} & \textbf{0.62}\\
  \end{tabular} 
  \caption{Estimates of $\beta_c$ for various choices of the fit parameters. In bold the best fits.}
\label{tab:betac}
\end{center}
  \end{table}
Another quantity easily accessible is the latent heat, this
quantity can be related to the height of the peak of the specific heat at the
critical temperature through:
\be
\frac{C_{L}(\beta_c(L))}{6L^4}=\frac{G^{2}}{4}+ \sum_{k=1}^{k_{max}}
C_k L^{-4k} 
\label{eq:peakscaling},
 \en
where $G$ is the latent heat.  The results for this observable are
reported in Tab.~\ref{tab:peak}. We fit the result with
Eq.~(\ref{eq:peakscaling}), see Tab.~\ref{tab:latentheat}. 

\begin{table}
\begin{center}
\begin{tabular}{c  c  c }
  L & $ \ \ \ C_V/(6V) $ peak present work \ \ \ & \ \ \ $ C_V/(6V) \
                                                   \ \ $ peak from \cite{Arnold:2000hf}  \\
  \hline
  8  & 0.000551(2) & 0.000554(1)  \\
  10 & 0.000384(2) & 0.000385(1)\\
  12 & 0.0002971(11) & 0.000298(1)\\
  14 & 0.0002537(8)& 0.000254(1)\\
  16 & 0.0002272(7)& 0.000226(2)\\
  18 & 0.0002097(5)& 0.000211(2)\\
  20 & 0.0002007(4)&  \\
  \end{tabular} 
  \caption{$C_V(\beta_c(L))$ evaluated with the LLR algorithm
    and reference data from \cite{Arnold:2000hf}. Results for a $20^4$
  lattice have never been reported before in the literature.}\label{tab:peak}
\end{center}
  \end{table}
\begin{table}
\begin{center}
\begin{tabular}{c  c  c c }
  $ L_{min} $ & $ k_{max} $ &  $ G $ & $\chi^{2}_{red}$  \\
  \hline
  14  & 1 & 0.02712(9) & 4.6  \\
  \hline
  12 & 1 & 0.0273(2) & 31\\
  \textbf{12} & \textbf{2} & \textbf{0.02688(7)} & \textbf{1.4}\\
  \hline
  10 & 1 & 0.0276(2)& 74 \\
  10 & 2 & 0.02710(12)& 9.7\\
  \textbf{10} & \textbf{3} & \textbf{0.02681(9)} & \textbf{1.4}\\
  \hline
  8 & 1 & 0.0281(4) & 335\\
  8 & 2 & 0.02731(15)& 26\\
  8 & 3 & 0.02703(11)& 6.7\\
  \end{tabular} 
  \caption{Estimates of $G$ for various choices of the fit parameters. In bold the best fits.}\label{tab:latentheat}
\end{center}
  \end{table}
The latent heat can be obtained also from the knowledge of the
location of the peaks of the probability density at
$\beta_c$ (of infinite volume), indeed in this case the latent heat is equal to energy
gap between the peaks. 
This direct measure can be used as crosscheck of the
 previous analysis. In the language of the density of
states the probability
density is simply given by
\be
P_\beta(E) = \frac{1}{Z} \rho(E) e^{\beta E}.
\en

\begin{figure}[t]
\begin{center}
\includegraphics[height=10cm]{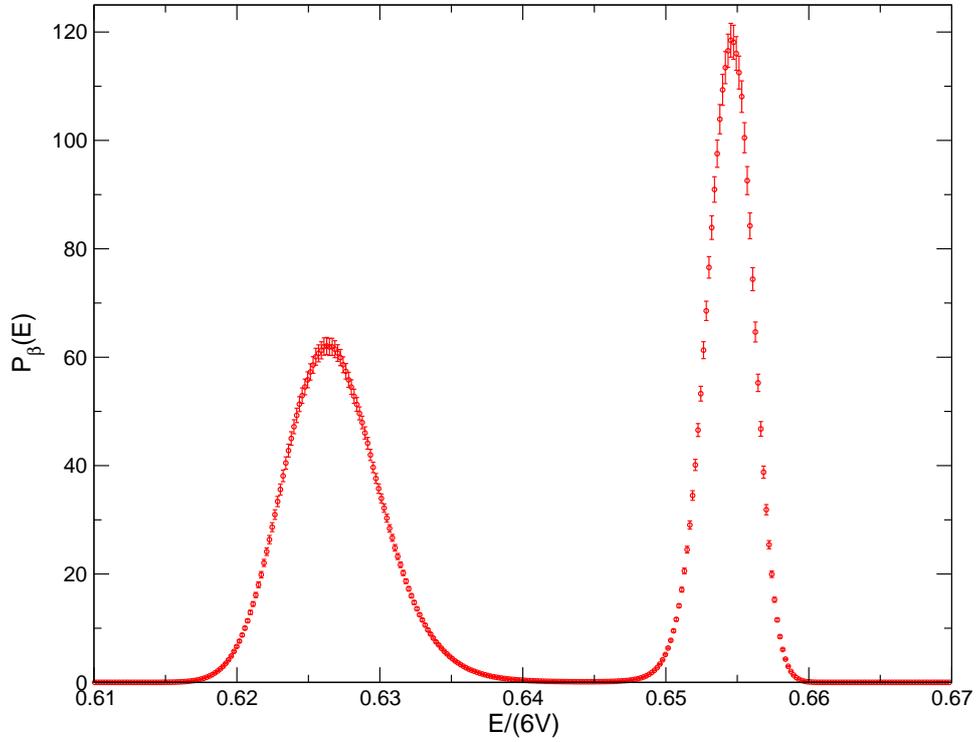} 
\end{center}
\caption{ Probability density for $L=20$ at $\beta_c$. The probability
is plotted at $\beta_c$ of infinite volume hence the peaks are not of
equal height.}
\label{fig:prob_dens}
\end{figure}

We have performed the study of the location in energy of the two peaks
of $P_{\beta_c}(E)$ and we have reported them in Tab.~\ref{tab:peakprob}. 
Also in this case we have performed a finite size scaling analysis to
extract the infinite volume behaviour:
\be
E_i(L)/(6V)=\epsilon_i +a_i e^{-b_i\,L}.
\en
A fit of the values in Tab.~\ref{tab:peakprob} yields
$\chi^{2}_{red,1}=0.67, \ \epsilon_1 =0.6279(9)$ and $\chi^{2}_{red,2}=0.2, \ \epsilon_2=0.65485(4)$.
The latent heat can be evaluated as $G=\epsilon_2-\epsilon_1=0.0270(9)$ which is in
perfect agreement with the estimates obtained by studying the
scaling of the specific heath.

\begin{table}
\begin{center}
\begin{tabular}{c  c  c }
  $ L $ & $ E_1/(6V) $ &  $E_2/(6V)$  \\
  \hline
  12  & 0.6263(5)  & 0.65580(14)  \\
  14  & 0.6264(2)  & 0.65532(5)  \\
  16  & 0.6272(2)  & 0.65512(4)  \\
  18  & 0.6274(4)  & 0.65495(6)  \\
  20  & 0.6275(2)  & 0.65491(7)  \\
  \end{tabular} 
  \caption{Location of the peak of the probability density in the two meta-stable phases.}\label{tab:peakprob}
\end{center}
  \end{table}

\subsection{Discretisation effects}
\label{sec:descaling}
\begin{figure}[htp] \centering{
\includegraphics[width=12cm]{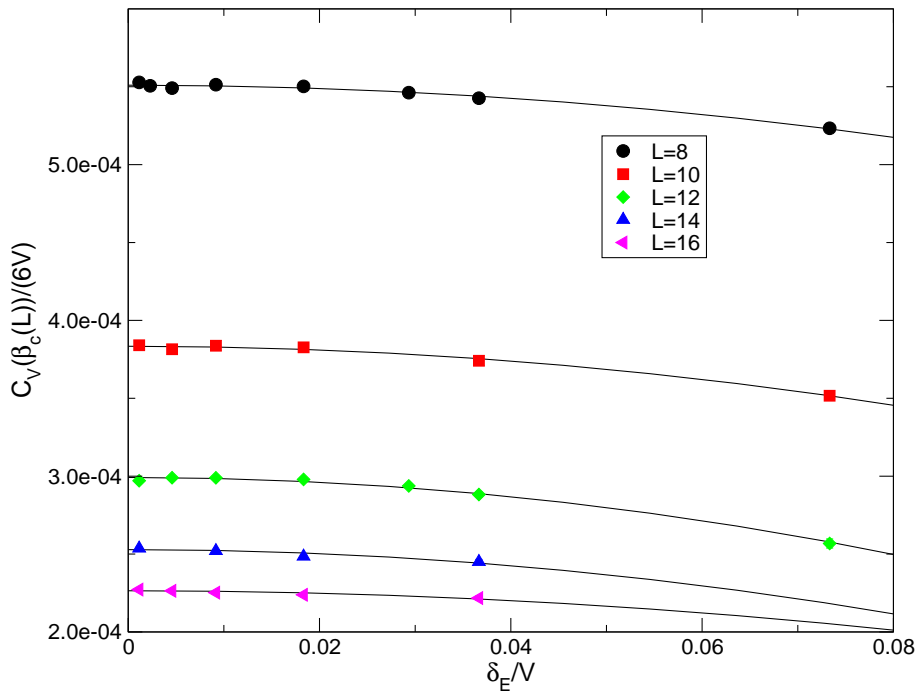}}
\caption{The peak of the $ C_{V}(\beta_C(L)) $  as function $\DE$.}
\label{fig:peakfigure}
\end{figure}

In this section we want to address the dependence of our observables
from the size of energy interval $\DE$. In order to quantify this
quantity we studies the dependence of the peak of the specific heat $C_{v,peak}$
with $\DE$ for various lattice sizes, namely $8,10,12,14,16$. 
In table \ref{tab:descaling} we report the lattice sizes and the corresponding
$\DE$ used to perform such investigation. For each pair of $\DE$
and volume reported we have repeated all our simulations and
analysis with the same simulation parameters reported in Tab.~\ref{tab:SimDet}.
\begin{table}[ht]
\begin{center}
\begin{tabular}{c  l}
  L & $(E_{max}-E_{min})/\DE$ \\[1mm]
  \hline
  10 & 8, 16, 32, 64, 128, 512\\
  12 & 8, 16, 20, 32, 64, 128, 512\\
  14 & 16, 32, 64, 512\\
  16 & 16, 32, 64, 128, 512\\
\end{tabular} 
\end{center}
\caption{Values of $\DE$ used to perform the study of the
  discretisation effects. The other simulation parameters are
 kept identical to the one reported in Tab. \ref{tab:SimDet}}
\label{tab:descaling}
\end{table}
The choice of the specific heat as an observable for such investigation
can be easily justified: we found that specific heath is much more
 sensible to the discretisation effects with respect to other simpler
 observables such as the plaquette expectation value. 
In Fig.~\ref{fig:peakfigure} we report an example of such study relative to $L=8$.
 We can confirm that all our data are scaling with 
 quadratic law in $\DE$ consistent with our findings in subsection
 \ref{sec:2.3}. 
Indeed by fitting our data with a form 
\be
C_V(\beta_c(L),\DE)=C_V(\beta_c(L)) +6 V b_{dis} \DE^{2},
\en
we found $\chi^{2}_{red} \sim 1 $ for all lattice sizes we investigated. 
We report in in Tab.(\ref{bdisc}) the values of $b_{dis}$.  Note that
the numerical values used in our finite size scaling analysis of the
peak of $C_V$ presented in the previous section are compatible with the results
extrapolated to $\DE = 0$ obtained here.


\begin{table}
\begin{center}
\begin{tabular}{c  c }
  $ L $ & $ b_{dis} $  \\
  \hline
  8  & -3.1(2) $10^{-10}$  \\
  10 & -5.9(4) $10^{-11}$   \\
  12  & -1.8(1)$ 10^{-11}$ \\
  14  & -4(1)$ 10^{-12}$\\
  16  & -9(3) $10^{-13}$\\
  \end{tabular} 
  \caption{The coefficient $b_{dis}$ for different lattice sizes.}\label{bdisc}
\end{center}
  \end{table}

\section{Discussion, conclusions and future plans}
\label{sect:conclusions}

The density of states $\rho (E)$ is a measure of the number of
configurations on the hyper-surface of a given action $E$. Knowing the
density of states relays the calculation of the partition function to
performing an ordinary integral. Wang-Landau type algorithms perform
Markov chain Monte-Carlo updates with respect to $\rho $ while
improving the estimate for $\rho $ during simulations. The LLR
approach, firstly introduced in~\cite{Langfeld:2012ah}, uses a
non-linear stochastic equation (see~(\ref{eq:kk11})) for this task and
is particularly suited for systems with continuous degrees of
freedom. To date, the LLR method has been applied to gauge
theories in several publications,
e.g.~\cite{Langfeld:2014nta,Langfeld:2013xbf,Pellegrini:2014gha,Gattringer:2015lra},
and has turned out in practice to be a reliable and robust method. In the
present paper, we have thoroughly investigated the foundations of the
method and have presented high-precision results for the U(1) gauge
theory to illustrate the excellent performance of the approach.

\medskip
Two key features of the LLR approach are:
\bi 
\item[(i)] It solves an {\it overlap} problem in the sense that the
  method can specifically target the action range that is of
  particular importance for an observable. This range might easily be
  outside the regime for which standard MC methods would not be
  able to produce statistics.
\item[(ii)] It features {\it exponential error} suppression: although
  the density of states $\rho $ spans many orders of magnitude, its
  linear approximation $\tilde{\rho}$ has a nearly-constant {\it
    relative} error (see subsection~\ref{sec:2.2}) and the numerical
  determination of $\tilde{\rho}$ preserves this level of accuracy.
\ei 
We point out that feature (i) is not exclusive of the LLR method, but
is quite generic for multi-canonical techniques~\cite{Berg:1992qua},
Wang-Landau type updates~\cite{Wang:2001ab} or hybrids
thereof~\cite{Berg:2006hh}.

\medskip
Key ingredient for the LLR approach is the double-bracket expectation
value~\cite{Langfeld:2012ah} (see (\ref{eq:kk7})). It appears as a
standard Monte-Carlo expectation value over a finite action interval
of size $\DE $ and with the density of states as a
re-weighting factor. The derivative of the density of states $a(E)$ emerges
from an iteration involving these Monte-Carlo expectation
values. This implies that their statistical error interferes with the
convergence of the iteration. This might introduce a bias preventing
the iteration to converge to the true derivative $a(E)$. We resolved
this issue by using the Robbins-Monro formalism~\cite{robbins1951}: 
we showed that a 
particular type of under-relaxation produces a normal distribution of
potential values $a(E)$ with the mean of this distribution coinciding
with the correct answer (see subsection~\ref{sec:2.2}). 

\medskip
In this paper, we also addressed two concerns, which were raised in
the wake of the publication ~\cite{Langfeld:2012ah}:
\bi
\item[(1)] The LLR simulations restrict the Monte-Carlo updates to a
  finite action interval and might therefore be prone to {\it
    ergodicity} violations.
\item[(2)] The LLR approach seems to be limited to the calculation of
  {\it action dependent observables only}.
\ei 
To address the first issue, we have proposed in subsections
\ref{sec:2.5} and \ref{sec:2.6} two procedures that are conceptually
free of ergodicity violations. The first method is based upon the
replica exchange method~\cite{Replica,Replicawl}: using overlapping
action ranges during the calculation of the double-bracket expectation
values offers the possibility to exchange the configurations of
neighbouring action intervals with appropriate probability (see
subsection~\ref{sec:2.5} for details). The second method is a standard
Monte-Carlo simulation but with the inverse of the estimated density
of states, i.e., $\tilde{\rho }^{-1}(E)$, as re-weighting factor. The
latter approach falls into the class of ergodic Monte-Carlo update
techniques and is not limited by a potential overlap problem: if the
estimate $\tilde{\rho }$ is close to the true density $\rho$, the
Monte-Carlo simulation is essentially a random walk in configuration
space sweeping the action range of interest.

To address issue (2), we firstly point out that the latter
re-weighting approach produces a sequence of gauge field
configurations that can be used to calculate {\it any} observable by
averaging with the correct weight. Secondly, we have developed in
subsection~\ref{sec:2.2} the formalism to calculate any observable by
a suitable sum over a combination of the density of states and
double-bracket expectation values involving the observable of
interest. We were able to show that the order of convergence (with the
size $\DE $ of the action interval) for these observables is the same
as for $\rho $ itself (i.e., ${\cal   O}(\DE^2)$).

\medskip
In view of the features of the density of states approach, our future
plans naturally involve investigations that either are enhanced 
by the  direct access to the partition function (such as the
calculation of thermodynamical quantities) or that are otherwise
hampered by the overlap problem. These, most notably, include complex
action systems such as cold and dense quantum matter. The LLR
method is very well equipped for this task since it is based upon
Monte-Carlo updates with respect to the positive (and real) estimate of
the density of states and features exponential error suppression which
might beat the resulting overlap problem. Indeed, a strong sign
problem was solved by LLR techniques using the original degrees of
freedom of the $Z_3$ spin
model~\cite{Langfeld:2014nta,Gattringer:2015lra}. We are currently 
extending these investigations to other finite density gauge
theories. QCD at finite densities for heavy quarks (HDQCD) is work in
progress.  We have plans to extend the studies to finite density 
QCD with moderate quark masses. 

\begin{acknowledgments}
We thank Ph.~de~Forcrand for discussions on the algorithm that lead
to the material reported in Subsect.~\ref{sec:2.6}. The numerical
computations have been carried out using resources
from HPC Wales (supported by the ERDF through the WEFO, which is part
of the Welsh Government)  and resources from the HPCC Plymouth. KL and
AR are supported by the Leverhulme Trust (grant RPG-2014-118) and STFC
(grant ST/L000350/1). BL is supported by STFC (grant ST/L000369/1). RP is supported by STFC (grant ST/L000458/1). 
\end{acknowledgments}

\appendix
\section{Reference scale and volume scaling}
\label{sect:appendix}

Here, we will present further details on the scaling of the density of
states $\rho (E)$ with the volume $V$ of our system. 
To this aim, we will work in the regime of a finite correlation length
$\xi $ such that the volume $V \gg \xi^4$. In the case of particle
physics, $\xi $ is a multiple of the inverse mass of the lightest
excitation of the theory. In this subsection, we do not address the
case of a correlation length comparable or larger than the size of the
system, as it might occur near a second order phase transition. 

\medskip 
Under these assumptions, the total action appears as a sum 
over uncorrelated contributions: 
\be 
E \; = \; \sum _{i=1}^v e_i \; , \hbo v = V/\xi ^4 \; , 
\label{eq:kkk1}
\en 
where the dimensionless variable $v$ is the volume in units of the
(physical) correlation length. To ease the notation, we will assume
that the densities $\rho $ and $\tilde{\rho }$ are normalised to one. 
Taking advantage of the above observation, we can introduce the
probability distribution $\tilde{\rho}(e_i)$ for the uncorrelated domains:
\be 
\rho (E)  =  \int \prod_{i=1}^v \dd e_i \; \delta\left( E \, - \, 
\sum _{k=1}^v e_k \right) \; \tilde{\rho }(e_1) \ldots \tilde{\rho }(e_v)
\; . 
\en 
Representing the $\delta $-function as Fourier integral, we find 
\bea 
\rho (E) &=& \int \frac{\dd \alpha }{2\pi} \; \int \prod_{i=1}^v \dd e_i
~ e ^{-i \,\alpha E } ~ e^{i  \,\alpha e_1} \ldots e^{ i \, \alpha
e_v}  ~ \tilde{\rho }(e_1) \ldots \tilde{\rho }(e_v)
\nonumber \\ 
&=&  \int \frac{\dd \alpha }{2\pi} \; e ^{-i\, \alpha E } \;
\Bigl\langle  e^{ i\,\alpha e } \Bigr\rangle ^v \; . 
\label{eq:kkk4}
\ena 
The latter equation is the starting point for a study of moments and
cumulants of the action expectation values and their scaling with the
volume. 

\medskip 
Cumulants of the action $E$ are defined by:
\bea 
( E^n)_c \; = \; (-i)^n \; \frac{d^n}{d \beta ^n} \, 
\ln \; \int dE \; \e ^{i \beta E} \; \rho (E) \; \Big\vert _{\beta=0}
\; . 
\label{eq:kkk5}
\ena 
Inserting (\ref{eq:kkk4}) into (\ref{eq:kkk5}), performing the $E$ and the
$\alpha $ integration leaves us with 
\be 
( E^n)_c \; = \; (-i)^n \; \frac{d^n}{d \beta ^n} \, 
\ln \; \Bigl\langle  \exp \{i \beta e \} \Bigr\rangle ^v
\Big\vert_{\beta=0} \; = \; v \;  ( e^n)_c \; , 
\label{eq:kkk7}
\en 
where the volume independent cumulants are defined by 
\be 
 ( e^n)_c \; = \;  (-i)^n \; \frac{d^n}{d \beta ^n} \, 
\ln \; \Bigl\langle  \exp \{i \beta e \} \Bigr\rangle 
\Big\vert_{\beta=0} \; . 
\label{eq:kkk8}
\en 
We here make the important observation that {\it all} cumulants are
proportional to the ``volume'' $v$ rather than powers of it. 
Re-summing (\ref{eq:kkk8}), i.e. using the identity
$$ 
\sum _n \frac{ i^n \alpha ^n}{n!} (e^n)_c \; = \; \ln \, 
\Bigl\langle  \exp \{i \alpha \,  e \} \Bigr\rangle \; , 
$$
we find for $\rho (E)$ in (\ref{eq:kkk4}) 
\be 
\rho (E) \; = \; \int \frac{d\alpha }{2\pi} \; \e ^{-i \alpha E } \;
\exp \Bigl\{ v \sum _n \frac{ (i \, \alpha )^n }{n!} \, (e^n)_c
\Bigr\} \; .
\label{eq:kkk30}
\en 
We perform the $\alpha $-integral by using the expansion 
\bea 
\rho (E) &=&  \exp \left\{ v \sum _{n=3}^\infty \frac{ (e^n)_c }{n!} \, 
\left( - \frac{d}{dE} \right) ^n \right\} \; \rho_0 (E) \; , 
\label{eq:31} \\
\rho _0 (E) &=& \int \frac{d\alpha }{2\pi} \; \e ^{-i \alpha E } \;
\exp \Bigl\{ i \, v (e)_c \, \alpha \; - \; v \, \frac{ (e^2)_c}{2} \;
\alpha ^2 \Bigr\} 
\nonumber \\ 
&=& \frac{1}{\sqrt{2 \pi v \, (e^2)_c}} \; 
\exp \left\{ - \frac{v}{2 \, (e^2)_c } \, [E/v -  (e)_c]^2 \,
\right\} \; . 
\label{eq:kkk32}
\ena 
In next-to-leading order, we obtain (up to an additive constant):
\be 
\frac{1}{v} \, \ln \rho(E) \; \approx \;  - \frac{[(e)_c - E/v]^2 }{2
 \, (e^2)_c } \; - \; \frac{ (e^3)_c }{ 6\, (e^2)_c^3 } 
[(e)_c - E/v ]^3 \; + \; {\cal O}(1/v) \; . 
\label{eq:kkk33}
\en
Hence, we find for the inverse temperature $a_k$
\be 
a_k \; = \; \frac{ d \, \ln \rho }{dE} \Big\vert_{E=E_k} \; \approx \; 
\frac{ (e)_c - E_k/v }{(e^2)_c } \; + \; \frac{(e^3)_c }{2 (e^2)^3_c }
\, \Bigl[(e)_c - E_k/v \Bigr]^2 \; . 
\label{eq:kkk34}
\en
We therefore confirm that $a_k$ is an intrinsic quantity, i.e., volume
independent. 
The curvature of $\ln \rho $ at $E=E_K$ is given by 
\be 
\frac{ d^2 \, \ln \rho }{dE^2} \Big\vert_{E=E_k} \; \approx \; 
- \, \frac{1}{v} \; \left[ \frac{1}{(e^2)_c } + \frac{(e^3)_c}{(e^2)_c^3} \, 
\Bigl[ (e)_c - E_k/v\Bigr] \right] \; . 
\label{eq:kkk35}
\en 
We therefore confirm the key thermodynamic assumptions in
(\ref{eq:kk2}) by explicit calculation: 
\be 
\frac{ d \, \ln \rho(E) }{ dE } \Big\vert _{E=E_k} \; = \; a_k \; = \;
{\cal O}(1) \;, \hbo 
\frac{ d^2 \, \ln \rho(E) }{ dE^2 } \Big\vert _{E=E_k} \; = \; 
{\cal O}(1/v) \; .
\label{eq:kkk36}
\en

\bibliography{density_long} 

\end{document}